%% file: girth.tex
\pdfoutput=1
\documentclass[11pt,letterpaper]{article}
\input{macros.tex}

\input{bib_macros}

\usepackage{url}

\def\cqedsymbol{\ifmmode$\lrcorner$\else{\unskip\nobreak\hfil
\penalty50\hskip1em\null\nobreak\hfil{$\lrcorner$}
\parfillskip=0pt\finalhyphendemerits=0\endgraf}\fi}

\usepackage{parskip}
\usepackage{microtype}

\newenvironment{recall}[1]{\smallskip
\noindent {\bf Reminder of #1  }\em}{}

\begin{document}

\begin{titlepage}
\def\thepage{}
\thispagestyle{empty}

\title{Approximating Cycles in Directed Graphs:\\ Fast Algorithms for Girth and Roundtrip Spanners}

\date{}
\author{
Jakub Pachocki
\thanks{
OpenAI,
\texttt{merettm@gmail.com}}
\hspace{0.5cm}
Liam Roditty
\thanks{
Bar Ilan University,
\texttt{liamr@macs.biu.ac.il}}
\hspace{0.5cm}
Aaron Sidford
\thanks{
Stanford University,
\texttt{sidford@stanford.edu}}
\and
Roei Tov
\thanks{
Bar Ilan University,
\texttt{roei81@gmail.com}}
\hspace{0.5cm}
Virginia Vassilevska Williams
\thanks{
MIT CSAIL,
\texttt{virgi@csail.mit.edu}}
}

\maketitle

\input{abstract}

\end{titlepage}

\input{intro}
\input{preliminaries}
\input{approach}

\input{covers}
\input{spanners}

\input{additive}
\input{lower}

\bibliographystyle{alpha}
\bibliography{girth2}

\begin{appendix}
\input{ball-app}
\input{sequential_clustering}
\end{appendix}

\end{document}

%% file: macros.tex
\usepackage{thmtools}
\usepackage{thm-restate}

\usepackage{mathrsfs}           
\usepackage{vmargin,fancyhdr}   
\usepackage{enumerate}

\usepackage{amsmath,amssymb}    
\usepackage{verbatim}           
\usepackage{xspace}             
\usepackage{graphicx,float}     
\usepackage{ifthen,calc}        
\usepackage{textcomp}           
\usepackage{fancybox}           
\usepackage{hhline}             
\usepackage{float}              

\usepackage[rflt]{floatflt}     
\usepackage[small,compact]{titlesec}
\usepackage{setspace}
\usepackage{subfigure}

\usepackage{cleveref}

\setpapersize{USletter}
\setmarginsrb{1in}{.5in}        
             {1in}{.5in}        
             {.25in}{.25in}     
             {.25in}{.5in}      
\newcommand{\showccc}[0]{0}
\newcommand{\ccc}[2][nothing]{
  \ifthenelse{\showccc=0}{}{
    \ensuremath{^{\Lsh\Rsh}}\marginpar{\raggedright\tiny\textsf{%
        \ifthenelse{\equal{#1}{nothing}}{}{\textbf{#1}\\}#2}}}}

\newcounter{hours}\newcounter{minutes}
\newcommand{\hhmm}{%
  \setcounter{hours}{\time/60}%
  \setcounter{minutes}{\time-\value{hours}*60}%
  \ifthenelse{\value{hours}<10}{0}{}\thehours:%
  \ifthenelse{\value{minutes}<10}{0}{}\theminutes}
\ifthenelse{\showccc=0}{\rhead{}}{\rhead{\today \ [\hhmm]}}
\newtheorem{theorem}{Theorem}[section]

\newtheorem{corollary}[theorem]{Corollary}
\newtheorem{definition}[theorem]{Definition}
\newtheorem{hypothesis}[theorem]{Hypothesis}

\newtheorem{lemma}[theorem]{Lemma}
\newtheorem{fact}[theorem]{Fact}

\newcommand{\Proof}[0]{\smallskip\noindent\textit{\textbf{Proof:}}\quad}

\newcommand{\Proofof}[1]{\smallskip\noindent\textit{\textbf{Proof of #1:}}\quad}

\newcommand{\QED}[0]{\hfill\ensuremath{\blacksquare}\medspace\\}

\newcommand{\R}[0]{\ensuremath{\mathbb{R}}}

\floatstyle{ruled}
\newfloat{algo}{tbp}{lop}
\floatname{algo}{Algorithm}

\usepackage{float}
\floatstyle{plain}\newfloat{myfig}{t}{figs}[section]
\floatname{myfig}{\textsc{Figure}}
\floatstyle{plain}\newfloat{myalg}{H}{algs}[section]
\floatname{myalg}{}
\newcommand{\poly}[1]{\text{poly}(#1)}

\newcommand{\eps}{\varepsilon}

\global\long\def\R{\mathbb{R}}

\newcommand{\bigO}{O}
\newcommand{\otilde}{\widetilde{\bigO}}
\global\long\def\Ot{\widetilde{\bigO}}

\global\long\def\defeq{\stackrel{\mathrm{{\scriptscriptstyle def}}}{=}}

\global\long\def\E{\mathbb{E}}

\usepackage{tikz}
\usetikzlibrary{positioning,chains,fit,shapes,calc}

\global\long\def\girth{\mathrm{girth}}
\global\long\def\outball{\mathrm{outball}}
\global\long\def\inball{\mathrm{inball}}
\global\long\def\Exp{\mathrm{Exp}}

%% file: bib_macros.tex
  \usepackage{nth}
  \usepackage{intcalc}

%% file: abstract.tex
\abstract{
	
	The {\em girth} of a graph, i.e. the length of its shortest cycle, is a fundamental graph parameter. Unfortunately all known algorithms for computing, even approximately, the girth and girth-related structures in directed weighted $m$-edge and $n$-node graphs require $\Omega(\min\{n^{\omega}, mn\})$ time (for $2\leq\omega<2.373$). In this paper, we drastically
	improve these runtimes as follows:
\\
	\begin{itemize}\setlength\itemsep{1em}
		\item \textbf{Multiplicative Approximations in Nearly Linear Time:} We give an algorithm that in $\otilde(m)$  time computes an $\otilde(1)$-multiplicative approximation of the girth as well as an $\otilde(1)$-multiplicative roundtrip spanner with $\otilde(n)$ edges with high probability (w.h.p).
		\item \textbf{Nearly Tight Additive Approximations:}
		For unweighted graphs and any $a \in (0,1)$ we give an algorithm that in $\otilde(mn^{1 - a})$  time  computes an $O(n^a)$-additive approximation of the girth, w.h.p.
		We show that the runtime of our algorithm cannot be significantly improved without a breakthrough in combinatorial boolean matrix multiplication. We also show that if the girth is $O(n^a)$, then the same guarantee can be achieved via a deterministic algorithm.
	\end{itemize}
	~\\Our main technical contribution to achieve these results is the first nearly linear time algorithm for computing roundtrip covers, a directed graph decomposition concept key to previous roundtrip spanner constructions. Previously it was not known how to compute these significantly faster than $\Omega(mn)$ time. Given the traditional difficulty in efficiently processing directed graphs, we hope our techniques may find further applications.

}

%% file: intro.tex

\section{Introduction}
\label{sec:intro}

%

The {\em girth} of a graph $G$ is the length of the shortest cycle in $G$. It is a natural and fundamental graph parameter that has been extensively studied (see Diestel's book~\cite{diestel} for a discussion) with research on its computation dating back to the 1960s. Perhaps, the most straightforward algorithm for the girth is simply to compute All-Pairs Shortest Paths (APSP). Surprisingly, this simple relationship leads to the best known algorithm for girth in a $n$-node graph with nonnegative weights: the breakthrough APSP algorithm of Williams~\cite{Williams14a} can compute the girth in $n^3/2^{\Theta(\sqrt{\log n})}$ time. For sparse weighted graphs with $m$ edges, an $O(mn)$ runtime was recently obtained by Orlin and Sede{\~{n}}o{-}Noda~\cite{orlin}, improving upon an $O(mn+n^2\log\log n)$ runtime that follows from using the best known sparse APSP algorithm~\cite{PettieR05}.

When the graph can be dense, all known girth algorithms run in $n^{3-o(1)}$ time (unless the weights are small integers bounded in absolute value by $M$ in which case an $\Ot(Mn^\omega)$ time is known~\cite{RodittyW11}). Vassilevska W. and Williams~\cite{WilliamsW10} explained this by showing that the girth in weighted graphs is equivalent to APSP in the sense that if one of the two problems has a ``truly subcubic'', $O(n^{3-\eps})$ time algorithm for some $\eps>0$, then both do. As it is a longstanding open problem whether APSP has a truly subcubic time algorithm, computing the girth in $O(n^{3-\eps})$ time would be a huge breakthrough.

One might wonder whether very sparse graphs allow for much faster than $O(mn)$ time girth algorithms (the above discussion says that dense graphs probably do not).
Very recently, Lincoln et al.~\cite{lincolnmn} showed that if there exists any integer $L$ such that for sparsity $m=\Theta(n^{1+1/L})$, one can 
obtain an $O(mn^{1-\eps})$ time algorithm for the girth of weighted directed graphs, then Max-Weight $k$-Clique would have surprisingly fast algorithms.
Weighted $k$-Clique is a notoriously difficult problem: it is the basis of several conditional lower bounds. Here, assuming its hardness implies that $mn^{1-o(1)}$ time is likely necessary for computing the girth exactly, for almost any sparsity. 


Because of the strong barriers for exact computation, efficient {\em approximation} algorithms are of interest.
Fast approximations for the girth in {\em undirected} graphs are possible and have been studied extensively. It is well known (\cite{althofer}) that for any integer $k\geq 1$, every weighted undirected $n$-node graph $G$ contains an $O(n^{1+1/k})$ edge $(2k-1)$-{\em spanner}, i.e. a subgraph that $(2k-1)$-approximates all pairwise distances in $G$.
Such $(2k-1)$-spanners can be computed in $\Ot(mn^{1/k})$ time~\cite{ThorupZ05}, and immediately imply efficient approximation algorithms for girth in undirected weighted graphs.
In unweighted undirected graphs even better results are known.
Itai and Rodeh~\cite{ItaiR78} gave an $O(n^2)$ time additive $1$-approximation algorithm, and follow-up work~\cite{RodittyW12,ll09} developed
even more efficient, combinatorial, truly subquadratic, approximation algorithms.	

Although impressive approximate girth algorithms are possible for undirected graphs, they all exploit properties particular to undirected graphs. In particular, not only do sparse spanners exist in undirected graphs, but it is also known~\cite{bondysim} that undirected graphs with $\Omega(n^{1+1/k})$ edges
must contain a $2k$-cycle. This fact is at the heart of obtaining fast girth approximation algorithms.

In contrast, directed graphs do not always contain sparse spanners and dense digraphs might not have any cycles at all (a directed bipartite clique is an example of each).
It is completely unclear what (if any) structure there is to exploit in directed graphs to obtain fast girth approximation algorithms.
Due to the close relationship between girth and APSP~\cite{WilliamsW10}, a-priori it could be that the girth problem in directed graphs suffers from the same problem as APSP in directed graphs and no finite approximation is even possible without resolving a major open problem about BMM.

A potential saving point is that while directed graphs may not contain sparse spanners, they do contain sparse {\em roundtrip spanners}. That is, if one uses $d(u,v)+d(v,u)$ for the distance between $u$ and $v$ instead of $d(u,v)$, then one can obtain a very similar result for directed graphs as in the undirected case: for all $k\geq 1$ and $\eps>0$, every $n$ node directed graph $G$ contains a $(2k+\eps)$-roundtrip spanner on $O(k^2/\eps n^{1+1/k}\log nW)$ edges (for nonnegative integer weights bounded by $W$). Furthermore, since roundtrip distances form a metric on the vertices, there are emulators for roundtrip distances 
with the same sparsity quality trade-offs as the best spanners (although they might be expensive to compute as computing the metric itself solves APSP). 

Unfortunately, while roundtrip spanners have been known to exist for over a decade (and roundtrip emulators for over 25 years~\cite{PelegS:89,PelegU:89,althofer} through the aforementioned reduction to spanners), the fastest algorithms for computing them run in $O(mn)$ time, essentially the time to solve APSP.

\subsection{Our Results}


In this paper we provide the first non-trivial approximation algorithms for computing the girth and related properties on directed graphs that run substantially faster than the roughly $\Omega(mn)$ time currently needed to solve APSP on a $n$-node, $m$-edge directed graph with non-negative weights.

We show how to compute $\otilde(1)$-multiplicative approximations to the \emph{girth} and construct \emph{multiplicative roundtrip} spanners in nearly linear time (See \Cref{sec:intro:mult_approx}), and we show how to compute additive approximations to the girth in time that is nearly tight under standard assumptions (See \Cref{sec:intro:add_approx}). To achieve these results we provide the first nearly linear time algorithms for computing roundtrip covers, a natural directed graph decomposition notion in prior work on roundtrip spanners \cite{CowenW04, RodittyTZ08} (See \Cref{sec:intro:techniques}).



\subsubsection{Multiplicative Approximations in Nearly Linear Time}
\label{sec:intro:mult_approx}

\cite{WilliamsW10} showed that the girth problem in weighted graphs is subcubically equivalent to APSP in general graphs.
Thus, obtaining a truly subcubic algorithm for girth in weighted graphs would imply a major breakthrough, and is a daunting task for current techniques.
In fact, {\em no} nontrivial combinatorial approximation algorithms were previously known even for the restricted case of unweighted directed graphs. On the other hand, we show how to obtain an $O(\log n)$ approximation to the girth in weighted graphs in slightly super-linear time (See \Cref{thm:girthapx}). Setting $k := \log n$ in this theorem allows us to compute an $O(\log^2 n)$ approximation to the girth in \emph{nearly linear} time.

\begin{restatable}[Multiplicative Girth Approximation]{theorem}{thmgirthapx}
	\label{thm:girthapx}
	For any $n$-node, $m$-node directed graph with nonnegative integer edge weights, with unknown girth $g$ and integer $k \geq 1$, in time $O(mn^{1/k}\log^{5} n)$ we can
	compute an estimate $\bar{g}$ such that
	$
	g\leq \bar{g}\leq O(k \log n)\cdot g
	$ with high probability.
\end{restatable}

Thus, by suffering only a logarithmic loss in the accuracy, one can obtain a girth estimate much faster than $mn$, a runtime that is likely optimal for exact computation.

%

Using our new directed graph decomposition algorithm we also show how to compute multiplicative roundtrip spanners in nearly linear time.
A {\em spanner} is a sparse subgraph that preserves the distances of the original graph with some multiplicative or additive approximation.
Since even preserving the asymmetric reachability structure of directed graphs may require $\Omega(n^2)$ edges (e.g. the complete directed bipartite graph), no sparse spanner yielding a finite multiplicative approximation is possible.
Instead, we consider spanners under the {\em roundtrip} distance metric, i.e. {\em roundtrip} spanners.

Given vertices $u$ and $v$ the {\em roundtrip} distance between $u$ and $v$ is the distance from $u$ to $v$ plus the distance from $v$ to $u$. Roundtrip distances were studied by Cowen and Wagner~\cite{CowenW04} in the context of routing. Later Roditty, Thorup and Zwick~\cite{RodittyTZ08} obtained roundtrip spanners for directed graphs that are almost as good in terms of their sparsity/approximation tradeoff as the spanners of undirected graphs~\cite{althofer}: $(2k-\eps)$-multiplicative approximation with $\tilde{O}((k^2/\eps) n^{1+1/k})$ edges for any integer $k\geq 1$ and $\eps\in (0,1)$. However, the construction of these spanners requires precomputing the roundtrip distances between \emph{all} pairs of vertices, resulting in a running time of roughly $\Omega(mn)$. We show how to nearly match this size/approximation tradeoff while decreasing the time needed to construct them to nearly linear in the number of edges in the graph.

\begin{theorem}[Multiplicative Roundtrip Spanners]  Given any $n$-node, $m$-edge directed graph with nonnegative integer edge weights and any $k\geq 1$ in time $O(mn^{1/k}\log^{5} n)$ we can compute an  $O(k\log n)$-multiplicative roundtrip spanner with $O(n^{1+1/k}\log^2 n)$ edges with high probability.
\end{theorem}

Our techniques are inherently parallelizable, and we provide the first work-efficient parallel algorithms for computing both the approximate girth and the strongly connected components of an unweighted directed graph with depth linear in the diameter of the computed objects (See \Cref{sec:parallel}).

\subsubsection{Nearly Tight Additive Approximations}
\label{sec:intro:add_approx}

Our techniques can also be used to obtain fast combinatorial algorithms that achieve {\em additive approximations} of the girth on unweighted graphs as follows. Let $a \in (0, 1)$ be any constant and suppose the girth $g$ is $<n^a/\log n$. Then, the algorithm from Theorem~\ref{thm:girthapx} will return w.h.p. in $\tilde{O}(m)$ time a cycle of length $O(n^a)$, which is (trivially) an additive $O(n^a)$ approximation of $g$. If on the other hand $g\geq n^a/\log n$, then if we take a random sample $S$ of $C n^{1-a}\log^2 n$ nodes for large enough constant $C$, then w.h.p. $S$ will contain a vertex of the shortest cycle. Then, running BFS from each node of $S$ will find the shortest cycle containing a node of $S$ and hence compute $g$ exactly:

\begin{corollary}[Additive Girth Approximation]\label{cor:add}
For any unweighted $n$-node, $m$-edge directed graph with unknown girth $g$ and $a \in (0,1)$ in time $\tilde{O}(mn^{1-a})$ we can
compute an estimate $\bar{g}$ such that
$
g\leq \bar{g}\leq g +  O(n^a)
$ with high probability.
\end{corollary}

Our algorithms for Theorem~\ref{thm:girthapx} and Corollary~\ref{cor:add} are combinatorial, but randomized. It is unclear whether they can be derandomized without incurring a large runtime cost. In particular, our algorithms use sampling to crudely estimate the sizes of reachability sets for all vertices in the graph. As far as we know, there are no faster deterministic ways to do this in the worst case than explicitly computing the reachability sets which requires $\Omega(\min\{n^\omega, mn\})$ time.
We partially derandomize Corollary~\ref{cor:add} using different techniques:


\begin{theorem}[Deterministic Additive Girth Approximation]
\label{thm:additve-appx} There is a {\em deterministic} combinatorial algorithm that for any unweighted $n$-node $m$-edge directed graph with unknown girth $g$ and parameters $a,\epsilon \in (0,1)$ computes in $\tilde{O}(\epsilon^{-2} mn^{1-a})$ time an estimate $\bar{g}$ such that $g \leq \bar{g} \leq g + O(n^\alpha)$ if $g \le n^a$ and $g \leq \bar{g} \leq (1+ \epsilon) g$  if $g > n^a$.
\end{theorem}

A natural question is whether the $\tilde{O}(mn^{1-a})$ runtime for $n^a$-additive approximation is necessary.
Surprisingly, we show that when it comes to combinatorial algorithms, Theorem~\ref{thm:additve-appx} and Corollary~\ref{cor:add} are optimal up to constant factors in the additive error, barring a breakthrough in BMM algorithms:

\begin{theorem}[Hardness for Improving Additive Running Time]
Suppose there is a combinatorial algorithm for some $\eps>0$ and $a=1/2$ that computes an additive $n^a-1$ approximation to the girth of any unweighted $n$-node $m$-edge directed graph in  $O(mn^{1-a-\eps})$ time. Then for some constant $\delta>0$ there is an $O(n^{3-\delta})$ time combinatorial algorithm for $n\times n$ BMM.
\end{theorem}

%
%
%
%
%
%
\subsection{Algorithmic Techniques : Roundtrip Covers in Nearly Linear Time}
\label{sec:intro:techniques}

Our key technical contribution towards achieving the majority of our algorithmic results is the first nearly linear time algorithm for computing \emph{roundtrip covers} of directed graphs. Informally, a roundtrip cover is a decomposition  of a directed graph into an overlapping collection of \emph{balls}, i.e. roundtrip distance induced subgraphs. It is required that the radius, or maximum roundtrip distance, in each ball be bounded and that any pair of vertices of bounded roundtrip distance appear together in some ball (See \Cref{sec:spanners} for formal definition). Computing such covers naturally yields multiplicative roundtrip spanners and has been considered in previous work \cite{CowenW04,RodittyTZ08}.

Unfortunately, 
all known roundtrip cover computation algorithms prior to this paper ran in at least $\Omega(mn)$ time. 
It was not clear how to efficiently manipulate the roundtrip metric for the purposes of computing such decompositions. It seemed that, in the worst case, one would have to compute almost the entire roundtrip metric explicitly, i.e. solve APSP.

We overcome this difficulty through a careful application of a few techniques. The first is natural: cluster the graph by growing balls of exponentially distributed radii or using exponential distribution based clustering techniques.  
(Similar ideas were used recently for parallel algorithms for undirected graph decomposition \cite{MillerPX13} and directed maximum flow \cite{EneMPS16}).
This does not distort too many roundtrip distances, but may fail to produce clusters of significantly smaller size.
The second is our key insight: we show that if we carefully seed such a clustering routine we can ensure that we either find a large cluster with small roundtrip diameter or we break the graph into significantly smaller pieces while sufficiently preserving roundtrip distances. Unfortunately, naively implementing such a procedure would be expensive (i.e. involve solving APSP). To circumvent this, we use another trick: a known sampling based approach to estimate the fraction of vertices that each vertex can reach or is reachable by within a given distance and show this suffices to pick seeds for clustering.

In short, we achieve our multiplicative approximations via a delicate combination of several powerful tools that have been defined and used before: (1) low diameter graph decompositions first introduced in \cite{Awerbuch85}, (2) using the exponential distribution for decomposition (e.g. in  
\cite{LinialS91, Bartal96, MillerPX13, EneMPS16}), 
(3) recursive graph decomposition  (e.g. in \cite{Bartal96, FRT, EneMPS16}), and (4) sampling based reachability set estimation (\cite{Cohen97}).
%
%
%
%
However, despite the prevalence of this machinery, it was an open question whether or not it could be leveraged to yield {\em any} running time improvement for the directed problems we consider. It was unclear a-priori if there was structure to exploit to quickly decompose the roundtrip metric and if the problems we consider were as hard as APSP.


Our key contribution is to show that this is not the case and there is in fact a way to rapidly reveal non-trivial directed graph structure sufficient to achieve our results. There are several pitfalls that occur when naively applying standard machinery to this problem and we believe the strength of our result is to show how to methodically overcome them (see \Cref{sec:approach:spanner}).
The lack of fast combinatorial primitives for directed graphs is occasionally referenced as indicative of the gap between recent progress on approximate undirected network optimization problems \cite{ChristianoKMST10, Madry10, LRS, Sherman13, KelnerLOS14, Peng14, Sherman16} and directed problems \cite{Madry13, LeeS14, CohenMSV16}. We hope our results and the insights that underly them may find future use.

\subsection{Additional Related Work}
For unweighted graphs, in the 1970s Itai and Rodeh~\cite{ItaiR78} showed that the girth can be computed in $O(nm)$ time via BFS, or in $O(n^{\omega})\leq O(n^{2.373})$-time using fast matrix multiplication~\cite{cw90,v12,legallmult}. These are still the best runtimes for the problem. Similarly to the relationship to APSP, \cite{WilliamsW10} showed that the girth in unweighted graphs is subcubically equivalent to Boolean Matrix Multiplication (BMM). A large open question in BMM is whether there exist truly subcubic ``combinatorial'' algorithms, that can avoid the sophisticated but often impractical tools for Strassen-like fast matrix multiplication (e.g. \cite{cw90,v12,legallmult}). The reduction from \cite{WilliamsW10} shows that either both BMM and girth have truly subcubic combinatorial algorithms, or neither of them does.

For the girth in undirected unweighted graphs, besides Itai and Rodeh's~\cite{ItaiR78} original $O(n^2)$ time additive $1$-approximation algorithm,
Roditty and Vassilevska W.~\cite{RodittyW12} presented an $\Ot(n^3/m)$-time additive $3$-approximation algorithm.
The additive $1$-approximation
 of \cite{ItaiR78} is also a multiplicative $4/3$-approximation. Lingas and Lundell~\cite{ll09} presented the first algorithm that breaks the quadratic time bound of \cite{ItaiR78}, at the price of a weaker approximation: their algorithm runs in $\Ot(n^{3/2})$ time and returns a multiplicative $8/3$-approximation. Roditty and Vassilevska W.~\cite{RodittyW12} presented an  $\Ot(n^{5/3})$-time deterministic multiplicative $2$-approximation algorithm. They also showed how to obtain a less than $2$ multiplicative approximation in truly subquadratic time for triangle-free graphs.

The history of combinatorial algorithms for BMM of $n\times n$ matrices is as follows.
Bansal and Williams~\cite{BansalW12} obtained an $O(n^3/\log^{2.25} n)$ time combinatorial algorithm improving on the $40$-year record of $O(n^3/\log^2 n)$ by the Four-Russians Algorithm~\cite{fourruss}.
The result of~\cite{BansalW12} was further improved by Chan~\cite{Chan15} to $O((n^3 /\log^{3} n) \log^{O(1)} \log n )$ time and most recently by Yu~\cite{Yu15} to $\hat{O}((n^3/\log^{4} n) \log^{O(1)} \log n )$.

Obtaining a truly subcubic time algorithm for APSP is among the most studied longstanding open problems in graph algorithms. In the 1970s Fredman~\cite{Fr76} showed that the $O(n^3)$ time classical Floyd-Warshall algorithm is not optimal by giving an $O(n^3 (\log\log n / \log n)^{1/3})$ time running time. Many polylogarithmic improvements followed, the last being $O(n^3 \log^3\log n/\log^2 n)$ by Chan~\cite{Chan07}. Two years ago, Williams~\cite{Williams14a} used techniques from circuit complexity, namely the polynomial method, to shave all polylogs, thus obtaining the current best bound for APSP, $n^3/2^{\Theta(\sqrt{\log n})}$.

Spanners in undirected weighted graphs were first studied by Awerbuch~\cite{Awerbuch85} and Peleg and Sch{\"{a}}ffer~\cite{PelegS89}.
Alth\"{o}fer et al.~\cite{althofer} showed that for every integer $k\geq 1$, every $n$-node graph, even if it is weighted, contains a multiplicative $(2k-1)$-spanner on $O(n^{1+1/k})$ edges. This result is optimal, conditioned on a well-known (and partially proven~\cite{WengerGraphs}) conjecture by Erd\"{o}s~\cite{ErdosExtG} about the existence of graphs of high girth.

\subsection{Organization}

The remainder of the paper is structure as follows. We introduce notation in \Cref{sec:preliminaries}, provide an overview of our approach in \Cref{sec:approach:spanner}, and show how to compute roundtrip covers in \Cref{sec:covers}. We then provide our algorithms for multiplicative approximations in \Cref{sec:spanners} and our algorithms for additive aproximations in \Cref{sec:additive}. We conclude with our lower bounds in \Cref{sec:lower}.

%% file: preliminaries.tex
\section{Preliminaries}
\label{sec:preliminaries}

Here we introduce various terminology we use throughout the paper.

\textbf{Graphs}: Throughout this paper we let  $G = (V, E, l)$ denote a directed graph with vertices $V$, edges $E \subseteq V \times V$, and non-negative edges lengths $l \in \R_{\geq 0}^{E}$. At times we consider unweighted graphs, that is graphs in which $l_e = 1$ for all $e \in E$ and in this case we will omit the $l$ altogether.

\textbf{Distances}: We let $d_G(u, v)$ denote the (shortest path) distance from $u$ to $v$ in $G$ and we abbreviate this as $d(u, v)$ when $G$ is clear from context. At times we consider shortest path distances over edge subgraphs $F \subseteq G$ and write $d_F(u, v)$ to denote the length of the shortest path from $u$ to $v$ using only the edges in $F$. In all cases we define $d(u, v) = \infty$ if there is not a path from $u$ to $v$.

\textbf{Roundtrip Spanners}: For $a, b \in V$ we refer to 
$d(a \rightleftarrows b) \defeq d(a, b) + d(b, a)$ as the roundtrip distance between $a$ and $b$. We call a subgraph $S \subseteq E$ an $\alpha$-multiplicative roundtrip spanner if
$
d_{S}(a \rightleftarrows b)\leq\alpha\cdot\left(d_{G}(a \rightleftarrows b)\right)
$ for all $a,b\in V$.


\textbf{Additive Approximation}: We call an estimate $\tilde{g}$ an $\alpha$-additive approximation to the girth $g$ of a directed graph if $g\leq \tilde{g}\leq g+\alpha$.

\textbf{Distance Measures}: For a directed graph $G = (V, E, l)$ we call $\min_{v \in V} \max_{v' \in V} d(v, v')$ the radius of $G$. We call $\max_{v, v' \in V} d(v, v')$ the diameter of $G$.

\textbf{Balls}: For a given metric, a \emph{ball} of radius $r$ around $v$ is the set of vertices within distance $r$ of $v$. We generally use the term `ball' to refer to balls in the roundtrip metric. For a directed graph $G$, we use $\inball_G(v, r)$ and $\outball_G(v, r)$ to denote the subsets of vertices of $G$ that can reach $v$ within distance $r$ or be reached from $v$ within distance $r$, respectively.

\textbf{Trees}: Given a directed graph $G = (V, E, l)$ we call a $T_{out} \subseteq E$ an out-tree with root $r \in V$ if the edges form an undirected tree and are all oriented away from $r$ (i.e. there is a $r$ to $v$ path for every node $v$ in the tree).  Similarly, we call $T_{in}$ an in-tree with root $r \in V$ if the edges form an undirected tree and are all oriented towards $r \in V$ (i.e. there is a $v$ to $r$ path for every node $v$ in the tree).

\textbf{Paths and Cycles}: A directed (simple) path $P=\langle u=v_1, v_2, \dots, v_k=v\rangle \subseteq V$ from $u$ to $v$ is an ordered set of vertices, where for every $i\in\{1,\dots, k-1\}$, $(v_i,v_{i+1})\in E$. A cycle $C=\langle u=v_1, v_2, \dots, v_k=v\rangle $ is a direct (simple) path with an additional requirement that $(v_k, v_1)\in E$.  If $P$ is a path and $u,v\in P$ such that $u$ precedes $v$ in $P$ then we denote by $P(u,v)$ the subpath of $P$ from $u$ to $v$. If $P_1$ and $P_2$ are paths in $G$, then we denote by $P_1 \cdot P_2$ the concatenation of $P_1$ and $P_2$.

\textbf{Running Times}: We use $\otilde$-notation to hide  logarithmic factors, i.e. $\otilde(f(n)) = O(f(n) \log^c f(n))$.

\textbf{Probability}: We use \emph{with high probability} (w.h.p)  to denote that an event happens with probability at least $1-1/O(\poly{n})$ where $n$ is the size of the input to the problem.

%% file: approach.tex
\section{Overview of the Approach}
\label{sec:approach:spanner}

Our approach for computing multiplicative roundtrip spanners is broadly inspired by the following simple general strategy for computing spanners in undirected unweighted graphs:

1. Repeat until there are no vertices left:
\begin{itemize}
	\item Grow a ball of random radius from a vertex.
	\item Add the edges in the computed shortest path tree to the spanner.
	\item Remove all vertices in the ball from the graph.
\end{itemize}
2. Recurse on the subgraph induced by the edges that have endpoints in different balls.

If the radii are chosen appropriately one can show that the shortest path trees approximately preserve the distance between the endpoints of all edges inside a ball and that not too many edges are cut (i.e. have endpoints in different balls). While there is a great body of work on efficiently constructing spanners with many desirable properties \cite{BaswanaTMP05, Cohen99, DorHZ00, ElkinP04, ThorupZ05}, this simple strategy suffices to provide a polylogarithmic multiplicative spanner in nearly linear time.

Unfortunately there are two serious issues that prevent us from easily extending this approach to compute roundtrip spanners for directed graphs:
\begin{itemize}
	\item Recursing on cut edges does not work (or even make sense).
	\item There may be problematic vertices that are at a small distance to
	(or from) all the vertices, but have a large roundtrip distance to every vertex.
\end{itemize}
We derive our algorithm by carefully addressing these two issues.

\subsection*{Issue \#1: How to Recurse?}

The first issue is immediate. If we grow multiple balls in a directed graph and attempt to recurse on cut edges, it may be the case that we disconnect the graph and the roundtrip distances for the cut edges become infinite.
Consequently, if we recurse on the cut edges alone we simply do not have enough information to recover the path information we lost. Therefore, it is not clear if there is any reasonable way to recurse on the edges that we may distort.

To alleviate this issue we instead build a randomized scheme where we reason directly about the probability of cutting or distorting  the roundtrip distance between any particular pair of vertices. Building on previous work on graph decomposition \cite{LinialS91, Bartal96, MillerPX13} and directed maximum flow \cite{EneMPS16}  we use the fact that if we grow a ball where the radii are chosen from an exponential distribution then we can directly reason about the probability that removing that ball cuts a cycle.
We then proceed to repeatedly grow balls of exponential radii, removing them in each iteration.
Since the exponential distribution is memoryless, we can show that the probability that this approach
cuts a cycle depends only on the parameter of the exponential distribution we use.

For completeness, in Appendix~\ref{sec:sequential} we prove formally that repeated exponential ball growing works. For our algorithms we instead use a slightly more sophisticated clustering technique described in Section~\ref{sec:clustering}.
This scheme works for similar reasons, but is more easily parallelizable. Ultimately, both techniques allow us to reason about the probability of destroying a cycle (rather than an edge) and repeat until all cycles corresponding to the roundtrip distances between vertex pairs are preserved with high probability. The difficulty remains in ensuring that we can actually terminate such a procedure in a small number of iterations (i.e. Issue \#2).

\subsection*{Issue \#2: How to Avoid Problematic Vertices?}

The second issue seems even more troubling. Suppose that there is a graph with non-trivial cycle structure we would like to approximate. To create a harder instance one could simply create a new graph by adding many new vertices each of which has one short length edge to every original vertex and one long length edge from every vertex. Clearly these new vertices do not affect the cycle structure of the original graph that we wish to approximate. However, any shortest path query from these new vertices will quickly explore the entire graph. Consequently, starting any sort of clustering from these vertices could be quite expensive in terms of running time, yet reveals very little information about the graph's cycle structure.

Even if we take a different approach and simply attempt to improve the running time of constructing the roundtrip spanners from prior work~\cite{RodittyTZ08}, a similar issue arises. Here the immediate issue is that the algorithm computes balls in the roundtrip metric. However, to do this, again we need to explore many vertices at a large distance in the roundtrip metric for analogous reasons.

To alleviate this issue, we use sampling to estimate, in near linear time, the fraction of vertices in $|V|$ of $O(r)$-balls around all nodes, up to a small additive error $\epsilon$.  This can be done in nearly linear time due to a clever technique of Cohen~\cite{Cohen97} (For completeness, see Appendix however, we give a self-contained construction tailored to our purposes in Section~\ref{sec:ball_size_estimation}.) This allows us to find the problematic vertices that can reach many vertices at distance $r$ yet are reachable by few at distance $r$ (or vice versa). By using this technique to find problematic vertices we can better bias the seeds of our decomposition routines and make more progress in nearly linear time. This is crucial to our algorithm.

\subsection*{Building the Algorithm}

Combining our ideas to deal with these two issues yields our algorithm. We estimate for every vertex the number of vertices at distance $O(r)$ both from and to it. In one case there is a vertex that can both reach and is reachable by many vertices. In this case, we can compute a large enough ball (in the roundtrip metric) that contains vertices of small roundtrip distance, and then we recurse on the rest of the vertices outside the ball.
In the other case, there are many vertices that either do not reach or are not reached by too many vertices at distance $O(r)$ and we can grow clusters from them and recurse on all the clusters. In either case we show that we do not need to recurse too many times and that ultimately, with constant probability (see Section~\ref{sec:pass}), any particular pair of vertices at small roundtrip distance is together in a cluster.
 Repeating this procedure multiple times yields our nearly linear time roundtrip cover algorithm (see Section~\ref{sec:covers}).

We use our roundtrip cover algorithm to compute multiplicative roundtrip spanners and obtain multiplicative estimates of the girth. A naive application of our procedure would yield a logarithmic dependence on the range of lengths in the graph. To avoid this, we show how to break a directed graph into smaller graphs reducing to subproblems where lengths vary only polynomially in the number vertices. Furthermore, we show that our algorithm is inherently parallel and we obtain new work / depth tradeoffs for these problems. As discussed, this also yields faster additive approximation for the girth, though new insights are needed to obtain deterministic results (see Section~\ref{sec:spanners}). As discussed in the introduction, our roundtrip cover algorithm is also used to compute additive approximations to the girth (see Section~\ref{sec:additive}).

%% file: covers.tex
\section{Roundtrip Covers}
\label{sec:covers}

\newcommand{\clusterin}{\textsc{Cluster-In}}
\newcommand{\clusterout}{\textsc{Cluster-Out}}
\newcommand{\clusterinout}{\textsc{Cluster-Out(-In)}}
\newcommand{\estimateballs}{\textsc{Estimate-Balls}}
\newcommand{\probcover}{\textsc{Probabilistic-Cover}}
\newcommand{\roundtripcover}{\textsc{Fast-Roundtrip-Cover}}

In this section we provide our main results on graph partitioning. In particular we show how to efficiently construct roundtrip covers, first introduced in \cite{RodittyTZ08}.

\begin{definition}[Roundtrip Covers, definition 2.4 in \cite{RodittyTZ08}]
    A collection $\mathcal{C}$ of balls is a \emph{$(k,R)$-roundtrip-cover} of a directed graph $G = (V, E, l)$ if and only if each ball in $\mathcal{C}$ is of radius at most $kR$, and for every $u, v \in V$ such that $d_G(u \rightleftarrows v) \leq R$, there is a ball $B \in \mathcal{C}$ such that $u, v \in B$.
\end{definition}

The main result of this section is the following theorem, stating that we can construct a $(O(k \log n), R)$-roundtrip-cover with high probability in $\otilde(mn^{1/k})$ time.

\begin{theorem}[Fast Roundtrip Cover]\label{thm:roundtrip-cover}
    The algorithm $\roundtripcover(G, k, R)$, returns a collection $\mathcal{C}$ that is an $(O(k \log n), R)$-roundtrip-cover of directed graph $G = (V, E, l)$, w.h.p. in time $O(mn^{1/k} \log^4 n)$.
	Moreover, every vertex $v \in V$ belongs to $O(n^{1/k} \log n)$ elements of $\mathcal{C}$.
\end{theorem}

Note that if $G$ has integer edge lengths between $0$ and $U$ we can immediately apply Theorem~\ref{thm:roundtrip-cover} for a value of $R$ that is a power of $2$ and obtain $O(k \log n)$-multiplicative roundtrip spanners with  $O(n^{1 + 1/k} \log^2 n \log U)$ edges in time $O(mn^{1/k} \log^4 n \log U)$ as well as compute an $O(k \log n)$ multiplicative approximation to the girth in the same running time. Consequently, proving Theorem~\ref{thm:roundtrip-cover} encapsulates much of the difficulty in achieving our desired algorithmic results. However, in Section~\ref{sec:spanners} we show how a more careful application of Theorem~\ref{thm:roundtrip-cover} yields even stronger results, completely removing the dependence on $U$.

The remainder of this section is dedicated to providing the algorithm  $\roundtripcover$ and proving Theorem~\ref{thm:roundtrip-cover}. First in Section~\ref{sec:clustering} we provide our main graph clustering tool, then in Section~\ref{sec:balls} we provide our technique for estimating the fraction of vertices reachable to and from each vertex at some radius. Finally, in Section~\ref{sec:pass} we put these tools together to provide $\roundtripcover$ and prove Theorem~\ref{thm:roundtrip-cover}.

%

\subsection{Clustering}
\label{sec:clustering}

Here we provide the primary clustering/partitioning technique we use for our algorithm. We provide an algorithm that partitions the vertices into regions of bounded radius of our choice centered around a chosen
subset of the vertices so that the probability of separating any two vertices of bounded roundtrip distance is small.
The ability to control the radii and choose the starting vertices is key to deriving our algorithm.

As discussed in the introduction we use an exponential-distribution-based clustering procedure so that we can argue directly about the probability of cutting any particular cycle. This allows us to apply this procedure multiple times and argue by union and Chernoff bounds that with high probability we do not cut any cycle that we want to approximate, and thus obtain a good approximation of any relevant cycle.
However, rather than simply growing balls of exponentially distributed radius and repeating (as discussed in Section~\ref{sec:approach:spanner}), we provide a different scheme in the flavor of \cite{MillerPX13, EneMPS16} that better parallelizes. For completeness we complement our analysis with a proof that this sequential ball growing scheme also works in Appendix~\ref{sec:sequential}.

\begin{figure}[ht]
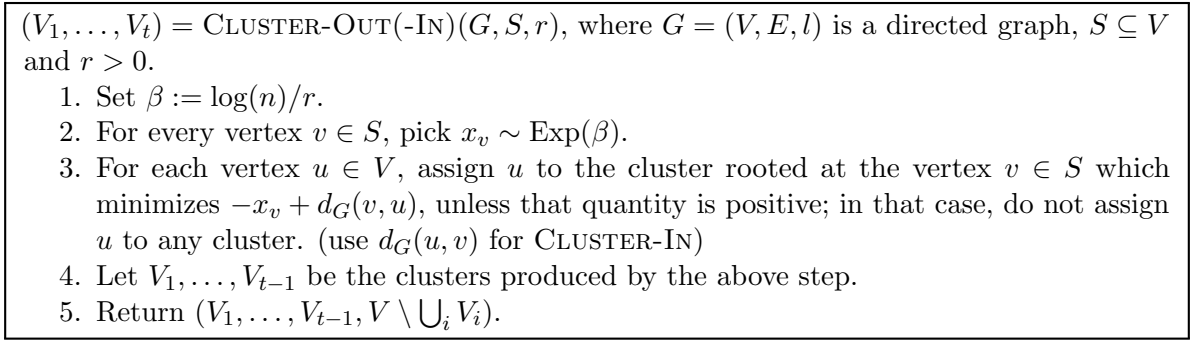

	\noindent
	\centering
	\fbox{
		\begin{minipage}{6in}
            \noindent $(V_1, \ldots, V_t) = \textsc{Cluster-Out(-In)} (G, S, r)$, where $G = (V, E, l)$ is a directed graph, $S \subseteq V$ and $r> 0$.
			\begin{enumerate}
                \item Set $\beta := \log(n) / r$.
                \item For every vertex $v \in S$, pick $x_v \sim \Exp(\beta)$.
                \item For each vertex $u \in V$, assign $u$ to the cluster rooted at the vertex $v \in S$ which minimizes $-x_v + d_G(v, u)$, unless that quantity is positive; in that case, do not assign $u$ to any cluster. (use $d_G(u, v)$ for $\textsc{Cluster-In}$)
                \item Let $V_1, \ldots, V_{t-1}$ be the clusters produced by the above step.
                \item Return $(V_1, \ldots, V_{t-1}, V \setminus \bigcup_i V_i)$.
			\end{enumerate}
		\end{minipage}
	}
	\caption{The clustering algorithm.}
	\label{fig:cluster-par}
\end{figure}

Our algorithms, $\clusterout$ and $\clusterin$ are given in Figure~\ref{fig:cluster-par}. Given a graph $G = (V, E, l)$, a set of vertices $S\subseteq V$ and a target radius $r$, the algorithm uses the exponential distribution to assign vertices in $G$ to clusters for each of the $v \in S$. The assignment is done in a way that ensures that these clusters each have bounded radius.
By our choice of assignment rule and distribution we formally show that the probability that two vertices of small roundtrip distance are not in the same cluster is sufficiently small.

In the remainder of this section we formally analyze this algorithm proving Lemma~\ref{lem:cluster}. The analysis we present is very similar to that of \cite{MillerPX13} and uses a subset of the ideas of \cite{EneMPS16}. The main difference is that we start only from a subset of the vertices $S$. Our analysis makes use of several facts regarding the exponential distribution which for completeness we prove in Appendix~\ref{sec:gen_tools}.

\begin{lemma}
\label{lem:cluster}
    Let $(V_1, V_2, \ldots, V_t)$ be the partition of $V$ returned by $\textsc{Cluster-Out}(G, S, r)$ (analogously of $\textsc{Cluster-In}$). Then, for any $c \geq 1$ we have
    \begin{enumerate}
        \item with probability at least $1 - n^{1 - c}$ for all
        $i < t$, the radius of the tree corresponding to $V_i$ is at most $c \cdot r$,
        \item for any pair of vertices $u, v$ at roundtrip distance at most $R$ in $G$, they are in the same set $V_i$ with probability at least
$
                \exp(-\log (n) R / r)
$.
    \end{enumerate}
Furthermore, the algorithm runs in time $O(m \log n)$.
\end{lemma}
\Proof
We prove the lemma for $\textsc{Cluster-Out}$ (the proof for $\textsc{Cluster-In}$ is analogous). We use various facts about the exponential distribution though this proof (See Section~\ref{sec:gen_tools} for their proof).
Note that the maximum radius of any cluster, $V_i$ is upper bounded by $\max_i x_i$ by design.
For every $i \in 1, \ldots, t$, we have
\[
    Pr\left[x_i \geq c \cdot r\right] \leq \exp(-c \cdot \beta r)
                                        = n^{-c} \,.
\]
By a union bound the maximum radius of is at most $c \cdot r$ with probability at least $1 - n^{1 - c}$.

To prove the
remainder of the lemma, fix $u, v \in V$ with roundtrip distance at most $R$ in $G$.
Assume $s \in S$ is the vertex minimizing $-x_s + \min(d_G(s, u), d_G(s, v))$, and that this quantity is less than $0$ (otherwise we have $u, v \in V_t$).
Let $T$ be the second smallest value of this quantity, or $0$, whichever is smaller.
Condition on the values of $x_s$ and $T$ and assume without loss of generality that $d_G(s, u) \leq d_G(s, v)$.
Then $u$ is assigned to the cluster rooted at $s$, and $u$ and $v$ can be separated only if $-x_s + d_G(s, v) > T$.
By the triangle inequality, this would imply $-x_s + d_G(s, u) + R > T$.
By assumption, we have $-x_s + d_G(s, u) < T$, or equivalently $x_s > d_G(s, u) - T$.
By the memoryless property of the exponential distribution (See Lemma~\ref{lem:exp-dist}), we see that the probability that the cluster rooted at $s$ contains both vertices $u$ and $v$ is at least
\begin{align*}
    \Pr\Big[x_s > d_G(s, u) - T + R \mid x_s > d_G(s, u) - T\Big]
=
	    \Pr\Big[\Exp(\beta)\geq R\Big]
=
		\exp(-\beta R),
\end{align*}
yielding the desired result.
\QED

\subsection{Estimating Ball Sizes}
\label{sec:balls}

To compute part of a roundtrip cover, ideally we would just partition the graph using the decomposition scheme of the previous graph and repeat until the clusters have good roundtrip diameter. Unfortunately, as discussed in Section~\ref{sec:approach:spanner} this approach fails as there may be problematic vertices that have a large low-radius ball in one direction, and a small low-radius ball in the other direction. In other words, calls to $\clusterin$ and $\clusterout$ with the wrong set $S$ might only yield trivial partitions, i.e. $V_1 = V$.

To alleviate this issue, we use a fast sampling approach to estimate the sizes of the $O(r)$-balls of all vertices, that allows us to identify these problematic vertices efficiently.
\begin{lemma}[\cite{Cohen97}]
\label{lem:balls-c}
For all $\epsilon \in (0,1)$ there is an algorithm $\textsc{Estimate-Balls}(G, r, \epsilon)$ that in  $O(m \epsilon^{-2} \log^2 n)$ time computes $n$-length vectors $s^{out}, s^{in}$, with the following property.
%
    For any vertex $u$, let $\bar{s}^{out}_u$ be the fraction of vertices in $V$ such that $d_G(u, v_i) \leq r$.
    Then, w.h.p., for all vertices $u$, $|\bar{s}^{out}_u - s^{out}_u| \leq \epsilon$, where $s^{out}_u$ is the component of $s^{out}$ corresponding to $u$.
    An analogous statement holds for $s^{in}$.
\end{lemma}

For completeness, we provide a self-contained proof of Lemma~\ref{lem:balls-c} in Appendix~\ref{sec:ball_size_estimation}.

\subsection{Fast Roundtrip Covers}
\label{sec:pass}

Combining the techniques of Section~\ref{sec:clustering} and Section~\ref{sec:balls} here we provide our efficient algorithm for constructing roundtrip covers, i.e. $\roundtripcover$, and prove the main theorem of this section, Theorem~\ref{thm:roundtrip-cover}, analyzing this algorithm.

We push much of the work of this algorithm to a subroutine $\probcover$ that performs the simpler task of constructing \emph{probabilistic} roundtrip cover: that is a partition of the vertex set such that any two vertices close enough in the roundtrip metric are in the same cluster with at least some fixed probability. Our main roundtrip cover construction is then simply a union of sufficiently many probabilistic roundtrip covers computed by $\probcover$.

\begin{figure}[ht]
	\noindent
	\centering
	\fbox{
		\begin{minipage}{6in}
            \noindent $\{B_1, B_2, \ldots\} = \probcover (G, r)$, where $G = (V, E, l)$ is a directed graph and $r> 0$.
			\begin{enumerate}
                \item Set $c > 1$ to a sufficiently large constant (for high probability bounds).
                \item If $V$ is empty, return $\emptyset$.
                \item Let $s^{out}, s^{in} := \textsc{Estimate-Balls}(G, c \cdot r, \frac{1}{8})$.
                \item Let $S^{out} := \{v \in V : s^{out}_v \geq \frac{3}{4}\}, S^{in} := \{v \in V : s^{in}_v \geq \frac{3}{4}\}$.
                \item If $S^{out} \cap S^{in} \neq \emptyset$:
                    \begin{enumerate}
                        \item Choose an arbitrary vertex $u \in S^{out} \cap S^{in}$.
                        \item (\emph{failure}) If $|\outball_G(u, c\cdot r) \cap \inball_G(u, c\cdot r)| < \frac{1}{4} \cdot |V|$, return $\{V\}$.
                        \item Pick $r_B$ uniformly at random in $[2c \cdot r, 2(c+1) \cdot r]$.
                        \item Let $B$ be a ball of radius $r_B$ in the roundtrip metric around $u$.
                        \item Let $G'$ be the graph induced by $G$ on $V\setminus B$.
                        \item Return $\{B\} \cup \probcover (G', r)$.
                    \end{enumerate}
                \item If $|S^{out}| \leq \frac{1}{2}\cdot|V|$:
                    \begin{enumerate}
                        \item Let $(V_1, \ldots, V_t) := \textsc{Cluster-Out}(G, V - S^{out}, r)$.
                    \end{enumerate}
                 Otherwise (we have $|S^{in}| \leq \frac{1}{2}\cdot|V|$):
                    \begin{enumerate}
                        \item[(b)] Let $(V_1, \ldots, V_t) := \textsc{Cluster-In}(G, V - S^{in}, r)$.
                    \end{enumerate}
                \item (\emph{failure}) If $\max_i |V_i| > \frac{7}{8} \cdot |V|$, return $\{V\}$.
                \item For $i = 1, \ldots, t$, let $G_i$ be the graph induced by $G$ on $V_i$.
                \item Return $\probcover (G_1, r) \cup \ldots \cup \probcover (G_t, r)$.
			\end{enumerate}
		\end{minipage}
	}
	\caption{Single pass of cover construction.}
	\label{fig:pass}
\end{figure}

The statement and analysis of $\probcover$ are the most technically involved results of this section. Our algorithm, $\probcover$, takes as input a directed graph $G$, a target radius $r$, and proceeds as follows. First  we use $\estimateballs$ from Section~\ref{sec:balls} to estimate the fraction of vertices in all balls of radius $O(r)$ up to an additive $1/8$. Then we consider two cases. In the first case we find that there is some vertex that can reach a large fraction of the vertices at distance $O(r)$ and can be reached by a large fraction of the vertices at distance $O(r)$. In this case we know that many vertices have a small roundtrip distance to this vertex so we simply output a roundtrip metric ball around this vertex and recurse on the remaining vertices. Otherwise, we know that there are many vertices that do not reach (or are not reachable by) many vertices at distance $O(r)$ and we can cluster to or from these vertices using $\clusterinout$ analyzed in Section~\ref{sec:clustering} and recurse on the clusters. In either case we recurse on subsets of vertices that are a constant fraction of the original size and hence only need to recurse a for a logarithmic number of iterations. We formally analyze this algorithm and prove that it has the desired properties in the following Lemma~\ref{lem:pass}.

\begin{lemma}
\label{lem:pass}
    Let $\mathcal{C} := \probcover (G, r)$. Then:
    \begin{enumerate}
        \item each $B \in \mathcal{C}$ is a ball of radius $O(r)$ in the roundtrip metric, w.h.p.,
        \item any pair of vertices $u, v$ at roundtrip distance at most $R$ in $G$ are in the same element of $\mathcal{C}$ with probability at least
            $\exp(-6 \log^2 (n) R / r)$, and
        \item every vertex $v \in V$ belongs to exactly one element of $\mathcal{C}$.
    \end{enumerate}
Furthermore, the algorithm runs in time $O(m\log^3 n)$.
\end{lemma}
\Proof
Property 3. is easily verified.

To prove property 1., first note that for large enough $c$ w.h.p. all calls to the subroutines $\textsc{Cluster-In}, \textsc{Cluster-Out}$ and $\textsc{Estimate-Balls}$ yield the guarantees described in \Cref{lem:cluster,lem:balls-c}.
Conditioning on this event, we show the failures in steps 5(b) and 7 of $\probcover$ never occur; this is enough to show the thesis.

First assume that there exists a vertex $u \in S^{out} \cap S^{in}$.
Then, by assumption, we have $|\outball_G(u, c\cdot r)| \geq (\frac{3}{4} - \frac{1}{8}) \cdot |V|\geq \frac{5}{8}\cdot |V|$ and $|\inball_G(u, c\cdot r)| \geq \frac{5}{8}\cdot |V|$.
Hence $|\outball_G(u, c\cdot r) \cap \inball_G(u, c\cdot r)| \geq \frac{1}{4} \cdot |V|$.
Hence the failure in step 5(b) cannot occur.
Thus, if the condition in step 5 holds, then whp the algorithm will return a cover.
Let's assume then that $S^{out} \cap S^{in}=\emptyset$. Then it must be that either $|S^{in}|\leq |V|/2$, or that $|S^{out}|\leq |V|/2$.
 Assume that $|S^{out}| \leq \frac{1}{2} \cdot |V|$ (the case $|S^{in}| \leq \frac{1}{2} \cdot |V|$ is analogous).
By assumption, the radii of all balls grown in calls to $\textsc{Cluster-Out}$ are at most $c\cdot r$, and so the sizes of the clusters constructed are at most $\frac{7}{8} \cdot |V|$ (by assumption on accuracy of $\textsc{Estimate-Balls}$).
The last cluster cannot be larger than $\frac{1}{2} \cdot |V|$ by construction.
Hence the failure in step 7 cannot occur.

To prove property 2., first note that in every recursive call, the size of the vertex set is multiplied by at most $7/8$.
Therefore, there are at most $\lceil \log_{8/7} n \rceil$ levels of recursion.
Now fix two vertices $u$ and $v$ at roundtrip distance at most $R \leq r$ in $G$.
In each level of recursion, the vertex set is partitioned by a call to $\textsc{Cluster-In}$ or $\textsc{Cluster-Out}$, or growing a roundtrip metric ball of radius chosen uniformly at random from $[2c\cdot r, 2(c+1)\cdot r]$.
For the first two cases, the probability $u$ and $v$ are not separated if they have not been separated previously is at least $\exp(-\log(n) R / r))$ by \Cref{lem:cluster}.
For the last case, the probability is easily seen to be at least $1 - R / (2\cdot r) \geq \exp(- \log(n) R / r)$.
Hence, the probability that $u$ and $v$ are not separated at all is at least
\begin{align*}
    \exp(-\log(n) R / r)^{\lceil \log_{8/7} n \rceil} &\geq \exp(- 6 \log^2 (n) R / r).
\end{align*}
Note that computing the ball in the roundtrip metric in step 5(d) reduces to two single source shortest path computations from $u$. Consequently, the running time is dominated by the $O(\log n)$ calls to $\textsc{Estimate-Balls}$.
\QED

\begin{figure}[ht]
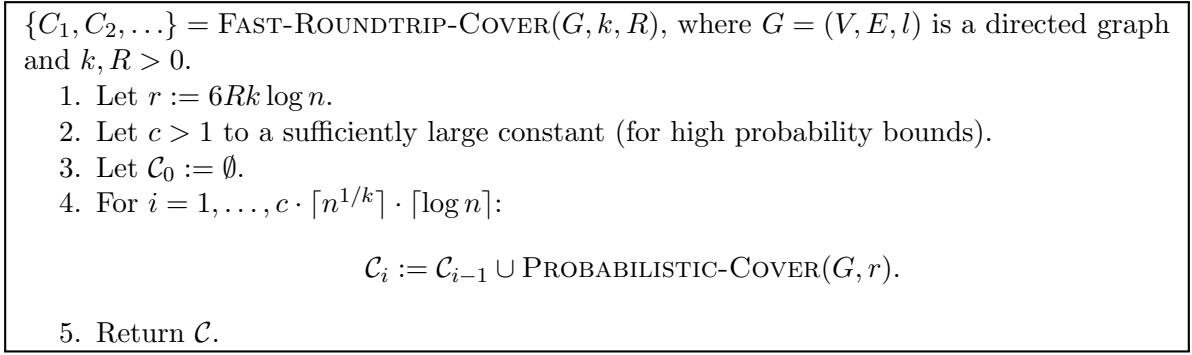

	\noindent
	\centering
	\fbox{
		\begin{minipage}{6in}
            \noindent $\{C_1, C_2, \ldots\} = \roundtripcover (G, k, R)$, where $G = (V, E, l)$ is a directed graph and $k, R > 0$.
			\begin{enumerate}
                \item Let $r := 6 R k \log n$.
                \item Let $c > 1$ to a sufficiently large constant (for high probability bounds).
                \item Let $\mathcal{C}_0 := \emptyset$.
                \item For $i = 1, \ldots, c \cdot \lceil n^{1/k} \rceil \cdot \lceil \log n \rceil$:
                \begin{align*}
                    \mathcal{C}_i := \mathcal{C}_{i - 1} \cup \probcover(G, r).
                \end{align*}
                \item Return $\mathcal{C}$.
			\end{enumerate}
		\end{minipage}
	}
	\caption{The fast roundtrip cover algorithm.}
	\label{fig:cover}
\end{figure}

With $\probcover$ in hand, we are ready to present the complete efficient algorithm for constructing roundtrip covers. The algorithm, $\roundtripcover$ is given in \Cref{fig:cover} and we conclude with its analysis, i.e. the proof of \Cref{thm:roundtrip-cover}.

\Proofof{\Cref{thm:roundtrip-cover}}
By \Cref{lem:pass}, w.h.p. all balls in $\mathcal{C}$ have the desired radius properties and the size assertions are easily verified.

It remains to show that w.h.p. any two vertices at short roundtrip distance share at least one cluster in $\mathcal{C}$.
Fix two vertices $u$ and $v$ at roundtrip distance at most $R$ in $G$.
By \Cref{lem:pass}, the probability they are in the same cluster in any single cover pass is at least $\exp(-6 \log^2 (n) R /r) = \exp(-\log(n) / k)$.
Hence, the probability they are separated in $\lceil n^{1/k} \rceil = \lceil \exp(\log(n) / k) \rceil$ independent passes is at most $\exp(-1)$.
Therefore, the probability they are separated in all the of $c \cdot \lceil n^{1/k} \rceil \cdot \lceil \log n \rceil$ passes is at most $\exp(-c\log n) = n^{-c}$.
\QED

%% file: spanners.tex
\section{Roundtrip Spanners and More}
\label{sec:spanners}

The analysis in Section~\ref{sec:covers} as encompassed by Theorem~\ref{thm:roundtrip-cover} yields our best result for unweighted graphs $G$; the union of $\textsc{Fast-Roundtrip-Cover}(G, k, R)$ for $R = 2^0, 2^1, \ldots, 2^{\lceil \log_2 n \rceil}$ is w.h.p. an $O(k \log n)$-multiplicative roundtrip-spanner of $G$.
More generally, for weighted graphs we obtain the following corollary:

\begin{corollary}
\label{cor:suboptimal-weight-spanner}
    Given a directed graph $G = (V, E, l)$ we can construct w.h.p. a $O(k \log n)$-roundtrip-spanner of $G$ with $O(n^{1 + 1/k} \log n \log (nW))$ edges in $O(mn^{1/k} \log^4 n \log (nW))$ time, where $W$ is the ratio between the largest and the smallest length in $G$.
\end{corollary}

The first aim of this section is to remove the dependence on $W$ from both the size of the spanner and the running time (\Cref{sec:weights}).
This allows us to prove the following result on spanner construction, and \Cref{thm:girthapx} as its corollary.

\begin{restatable}{theorem}{thmfastspanner}
\label{thm:fastspanner}
    The algorithm $\textsc{Fast-Roundtrip-Spanner}(G, k)$ in $O(mn^{1/k} \log^4 n)$ time computes w.h.p. an $O(k \log n)$-roundtrip-spanner of $G$ of size $O(n^{1+1/k} \log^2 n)$.
\end{restatable}

Then we shall investigate parallel algorithms that result from our scheme, obtaining the following results (\Cref{sec:parallel}).

\begin{restatable}{theorem}{thmscc}
\label{thm:scc}
    Given an unweighted directed graph $G$ and an upper bound $R$ on the maximum diameter of any strongly connected component of $G$, we can w.h.p. compute the strongly connected components of $G$ in $\otilde(m)$ work and $\otilde(R)$ depth.
\end{restatable}

\begin{restatable}{theorem}{thmparallelgirth}
\label{thm:parallelgirth}
    Given an unweighted directed graph $G$, we can w.h.p. compute an $O(k \log n)$ approximation to the girth of $G$ in $\otilde(mn^{1/k})$ work and $\otilde(\girth(G))$ depth.
\end{restatable}

\subsection{Removing the Dependence on Edge Lengths}
\label{sec:weights}

Our algorithm for constructing the spanner will remain based on the idea of taking a union of $(O(k \log n), R)$-roundtrip-covers over $R \in \mathcal{R}$, for some set $\mathcal{R}$ such that every roundtrip distance in $G$ is a constant factor smaller than some element of $\mathcal{R}$.

The main idea in removing the dependence on the lengths of the edges is that for a fixed value of $R$, we do not have to consider all the edges in $G$ when constructing a $(O(k \log n), R)$-roundtrip-cover.
First, note that we can remove all the edges longer than $R$, as that does not change any roundtrip distance smaller than $R$.
Simultaneously, for any strongly connected component of edges shorter than $R / n$, we can replace it by a single vertex.
Indeed, uncontracting all such vertices after obtaining a roundtrip cover will increase the length of any path found by only an additive $R$.
Finally, we can remove all the edges that do not participate in any strongly connected component, as that does not impact any roundtrip distances.
The idea is similar to those given in \cite{CohenMPPX14, EneMPS16}; the main difference from the scheme of \cite{EneMPS16} is in preserving only edges that are parts of connected components of edges shorter than $R$.
The described process is formalized in \Cref{def:collapse}.

\begin{definition}
\label{def:collapse}
    Let $G = (V, E, l)$ be a directed graph and $x_L, x_R \in \mathbb{R}$ be such that $0 < x_L < x_R$.
    We construct $G$ \emph{collapsed} to $[x_L, x_R]$ by:
    \begin{itemize}
        \item merging any vertices that can reach each other while following only edges of length at most $x_L$,
        \item removing all edges longer than $x_R$,
        \item removing all edges whose endpoints cannot reach each other while following only edges of length at most $x_R$, and
        \item removing all vertices of degree $0$ remaining afer the above operations.
    \end{itemize}
\end{definition}

To simplify notation, we define the $L_\infty$-roundtrip distance $d^\infty_G(u, v)$:

\begin{definition}
    For a directed graph $G = (V, E, l)$ and a pair of vertices $u, v \in V$, we define the $L_\infty$-roundtrip distance between $u$ and $v$, denoted $d^\infty_G(u, v)$, as the minimum value of $d$ such that there is a cycle $C$ containing $u$ and $v$ such that $l_e \leq d$ for all $e \in C$.
\end{definition}

We now show that performing the process for all $R \in \{2^t : t \in \mathbb{Z}\}$ results in a collection of graphs with a bounded size.

\begin{lemma}
\label{lem:gt}
    Let $G = (V, E, l)$ be a directed graph.
    For every $t \in \mathbb{Z}$, let $G^{(t)}$ be $G$ collapsed to $[2^t / n, 2^t]$.
    Then the total number of edges in all $G^{(t)}$ is $O(m \log n)$, and the total number of vertices in all $G^{(t)}$ is $O(n \log n)$.
\end{lemma}
\Proof
    Fix an edge $e = (u, v) \in E$ and $t$ such that $e$ is an edge in $G^{(t)}$.
    Note that since $e$ is not contracted, we must have $d^\infty_G(u, v) > 2^t / n$.
    Simultaneously, since $e$ is part of a strongly connected component in $G$ consisting of edges of length at most $2^t$, we must have $d^\infty_G(u, v) \leq 2^t$.
    Hence $t - \log_2 d^\infty_G(u, v) \in [0, \log_2 n)$.
    Therefore $e$ is included in $O(\log n)$ of the graphs $G^{(t)}$.

    Assume that $u$ is a vertex of $G^{(t)}$.
    By construction, it must be part of a nontrivial strongly connected component in $G^{(t)}$, and so it is merged with another vertex in $G^{(t')}$ for all $t' \geq t + \log_2 n$.
    Since there are only $O(n)$ possible vertices that can result from merging vertices in $V$, and each of them appears in $O(\log n)$ graphs $G^{(t)}$, we obtain the thesis.
\QED

If we can construct all the graphs $G^{(t)}$ efficiently, we can simply run $\textsc{Fast-Roundtrip-Cover}$ on each of them to obtain a spanner for $G$.
Following the idea of the proof of \Cref{lem:gt}, to construct all of $G^{(t)}$, is enough to compute for each edge $(u,v) \in E$ the value of $d^\infty_G(u, v)$.
This is obtained by the algorithms $\textsc{Roundtrip-$L_\infty$-Spanner}$ and $\textsc{Find-Collapse-Times}$, described below.
The algorithm $\textsc{Find-Collapse-Times}$ computes $d^\infty_G(u, v)$ for every edge $(u, v)$ in $G$, assuming that all the edges have distinct weights from $1$ to $m$.

\begin{figure}[ht]
	\noindent
	\centering
	\fbox{
		\begin{minipage}{6in}
            \noindent $s = \textsc{Find-Collapse-Times} (G, (e_L, \ldots, e_R))$, where $G = (V, E)$ is a directed graph, $L, R \in \mathbb{N}$ with $1 \leq L \leq R$.
			\begin{enumerate}
                \item If $L = R$, set $s_e = L$ for all $e \in E$ and return $s$.
                \item Let $M = \lfloor (L + R) / 2 \rfloor$.
                \item Let
                    \begin{align*}
                    E' := \{e \in E | e \mbox{ is contained inside a SCC of the graph } (V, \{e_L, \ldots, e_M\})\}.
                    \end{align*}
                \item Let $V'$ be $V$ with edges in $E'$ contracted.
                \item Let $s' := \textsc{Find-Collapse-Times}((V, E'), (e_L, \ldots, e_M))$.
                \item Let $s'' := \textsc{Find-Collapse-Times}((V', E \setminus E'), (e_{M+1}, \ldots, e_R))$.
                \item Return $s'$ merged with $s''$.
			\end{enumerate}
		\end{minipage}
	}
	\caption{The recursive algorithm for computing collapse times.}
	\label{fig:collapse}
\end{figure}

\begin{lemma}
\label{lem:times}
    Let $G = (V, E)$ be a directed graph and $(e_L, \ldots, e_R)$ be a sequence of edges on $V$.
    Assume that every edge in $E$ is contained in a strongly connected component of $(V, \{e_L, \ldots, e_R\})$.
    Let $s = \textsc{Find-Collapse-Times}(G, (e_L, \ldots, e_R))$.
    Then for every $e \in E$, it holds that $s_e$ is the minimum $i$ such that $e$ is contained in a strongly connected component of $(V, \{e_L, \ldots, e_{s_e}\})$.
    Moreover, the algorithm runs in $O((|V| + |E| + (R - L + 1)) \log (R - L + 1))$ time.
\end{lemma}
\Proof
    Correctness is easily proven by induction.
    To bound the running time, it is enough to observe that every recursive call halves $(R - L + 1)$, and every edge in $E$ is only passed to one recursive call.
\QED

The algorithm $\textsc{Roundtrip-$L_\infty$-Spanner}$ constructs an $O(n)$-sized subset $F$ of the edges of $G$ that preserves the $L_\infty$-roundtrip distances between vertices of $G$.
It also returns a tree $T$ containing all the vertices that can result from collapsing cycles of maximum edge length lower than some bound, with edges of $T$ describing the hierarchical structure on them.
Lowest common ancestor queries on $T$ enable us to efficiently compute $d^\infty_G(u, v)$ for any $u, v$.

\begin{figure}[ht]
	\noindent
	\centering
	\fbox{
		\begin{minipage}{6in}
            \noindent $(F, T) = \textsc{Roundtrip-$L_\infty$-Spanner} (G)$, where $G = (V, E, l)$ is a directed graph.
			\begin{enumerate}
                \item Remove from $G$ any edges that are not part of a strongly connected component.
                \item Let $e_1, \ldots, e_m$ be the edges of $G$, ordered by increasing length.
                \item Let $s := \textsc{Find-Collapse-Times}(G, (e_1, \ldots, e_m))$.
                \item Let $V_0 = V, F_0 = \emptyset$.
                \item Let $T = (V, \emptyset)$.
                \item For $i$ in $1, \ldots, m$:
                    \begin{enumerate}
                        \item Let $E_i$ be the set of edges $e$ for which $s_e = i$.
                        \item Let $E_i'$ be union of any out- and in-trees for the strongly connected components of $(V_{i-1}, E_i)$.
                        \item Let $F_i := F_{i-1} \cup E_i'$.
                        \item Let $V_i$ be the set of vertices obtained from $V_{i-1}$ by contracting all of $E_i$ (equivalently $E_i'$).
                        \item Label every vertex of $V_i \setminus V_{i-1}$ by $l(e_i)$.
                        \item Add all vertices of $V_i \setminus V_{i-1}$ to $T$.
                        \item For every vertex $u \in V_{i-1}$ that was contracted into a vertex $v \in V_i\setminus V_{i-1}$, add an edge between $v$ to $u$ to $T$.
                    \end{enumerate}
                \item Return $(F_m, T)$.
			\end{enumerate}
		\end{minipage}
	}
	\caption{The algorithm for computing a small subset of edges that preserves $L_{\infty}$-roundtrip distance.}
	\label{fig:linfspanner}
\end{figure}

\begin{lemma}
\label{lem:linf}
    Let $G = (V, E, l)$ be a directed graph.
    Let $(F, T) = \textsc{Roundtrip-$L_\infty$-Spanner}(G)$.
    Then:
    \begin{enumerate}
        \item $F \subseteq E$ is such that for any pair of vertices $u, v$ contained in a cycle in $G$ with maximum edge length $R$, there exists a cycle in $(V, F)$ containing $u$ and $v$ with maximum edge length $R$,
        \item $|F| = O(n)$, and
        \item for any two vertices $u, v \in V$, the label of the lowest common ancestor of $u$ and $v$ in $T$ is equal to $d^\infty_G(u, v)$.
    \end{enumerate}
    Moreover, the algorithm works in $O(m \log n)$ time.
\end{lemma}
\Proof
By $\Cref{lem:times}$, the application of $\textsc{Find-Collapse-Times}$ computes for each edge $(u, v) \in E$ the value of $d^\infty_G(u, v)$.
    The claims of the lemma follow by construction.
\QED

Finally, we describe our complete algorithm for computing roundtrip distance spanners of weighted graphs.
The algorithm first computes all graphs $G^{(t)}$ using a call to $\textsc{Roundtrip-$L_\infty$-Spanner}$ and the ideas of the proof of \Cref{lem:gt}.
It then invokes $\textsc{Fast-Roundtrip-Cover}$ for each of $G^{(t)}$ and returns the union of the results, together with a $L_\infty$-roundtrip distance spanner for $G$ to account for the collapsed clusters in $G^{(t)}$.

\begin{figure}[ht]
	\noindent
	\centering
	\fbox{
		\begin{minipage}{6in}
            \noindent $F = \textsc{Fast-Roundtrip-Spanner} (G, k)$, where $G = (V, E, l)$ is a directed graph and $k \geq 1$.
			\begin{enumerate}
                \item Let $(F_0, T) := \textsc{Roundtrip-$L_\infty$-Spanner}(G)$.
                \item For all $t \in \mathbb{Z}$, let $G^{(t)}$ be $G$ collapsed to $[2^t / n, 2^t]$.
                \item Let $i := 0$.
                \item For every $t$ such that $G^{(t)}$ is nonempty:
                \begin{enumerate}
                    \item $\mathcal{C}_i := \textsc{Fast-Roundtrip-Cover}(G^{(t)}, k, 2^t)$.
                    \item $F_{i + 1} := F_i \cup \mbox{ shortest path trees to and from roots of each ball in $\mathcal{C}_i$} $.
                    \item $i := i + 1$.
                \end{enumerate}
                \item Return $F_i$.
			\end{enumerate}
		\end{minipage}
	}
	\caption{The roundtrip spanner algorithm.}
	\label{fig:spanner}
\end{figure}

\thmfastspanner*

\Proofof{\Cref{thm:fastspanner}}
    First note that for each $t$, $F_0$ provides a low-cost spanner for every collapsed vertex of $G^{(t)}$.
    By uncollapsing the collapsed vertices of $G^{(t)}$ and adding in edges from $F_0$, the length of any path in the roundtrip cover at radius $2^t$ increases by at most an additive $2^t$.
    Since edges larger than $2^t$ have no influence on roundtrip distances not larger than $2^t$, we see that the roundtrip covers computed for each $G^{(t)}$ are also roundtrip covers for $G$ (after adding the edges of $F_0$).

    To obtain the claimed running time, we need to show that the nonempty graphs $G^{(t)}$ can be computed efficiently.
    Following the idea of the proof of \Cref{lem:gt}, we see that it is sufficient to compute for every edge $(u, v)$ the value of $d^\infty_G(u, v)$.
    By \Cref{lem:linf}, this is easily done using lowest common ancestor queries on $T$.
\QED

\Cref{thm:girthapx} is an immediate corollary of \Cref{thm:fastspanner}.

\thmgirthapx*

\Proofof{\Cref{thm:girthapx}}
    It is sufficient to execute $\textsc{Fast-Roundtrip-Spanner}(G, k)$.
    The smallest diameter of any cluster computed in calls to $\textsc{Fast-Roundtrip-Cover}$ will be no larger than $O(k \log n)\cdot \girth(G)$.
\QED

\subsection{Parallel Strongly Connected Components and Girth Estimation}
\label{sec:parallel}

Our main subroutine, $\textsc{Fast-Roundtrip-Cover}$, is inherently parallelizable.
This enables us to obtain a new parallel algorithm for computing strongly connected components in nearly linear work, and depth proportional to the maximum diameter of a strongly connected component (assuming access to a known upper bound).
To our knowledge, no previous guarantees of this type have been known, despite the classical status of analogous guarantees for problems such as parallel $u$-$v$ reachability in directed graphs.

\thmscc*

To prove this result, we first formally state the parallel runtime guarantees of $\textsc{Fast-Roundtrip-Cover}$.

\begin{lemma}
\label{lem:parallel-cover}
    For unweighted graphs, a parallel version of $\textsc{Fast-Roundtrip-Cover}(G, k, R)$ can be implemented to work in $\otilde(m n^{1/k})$ work and $\otilde(R)$ depth.
\end{lemma}
\Proof
    Since $\textsc{Probabilistic-Roundtrip-Cover}$ has only $O(\log n)$ levels of recursion, and the separate calls to it can be made in parallel, the bottleneck for depth is computing shortest paths.
    Since for unweighted graphs any paths computed in calls to $\textsc{Estimate-Balls}$, $\textsc{Cluster-Out}$ and $\textsc{Cluster-In}$ are of length $\otilde(R)$, the thesis follows by employing standard parallel breadth first search (cf. \cite{MillerPX13}).
\QED

We now proceed to prove \Cref{thm:scc}.

\Proofof{\Cref{thm:scc}}
    We start by computing $\mathcal{C} := \textsc{Fast-Roundtrip-Cover}(G, \log n, R)$.
    Now note that w.h.p., for any pair of vertices $u$ and $v$, they are part of the same cluster in $\mathcal{C}$ if and only if they are in the same strongly connected component of $G$.
    Hence, it is enough to compute weakly connected components of the relation of being part of the same cluster in $\mathcal{C}$; this is achieved with classical parallel algorithms \cite{ShiloachV82, ReifS92}.
\QED

Another corollary
is that we can parallelize our girth estimation algorithm for unweighted graphs.

\thmparallelgirth*

\Proofof{\Cref{thm:parallelgirth}}
    It suffices to invoke $\textsc{Fast-Roundtrip-Cover}(G, k, R)$ for $R \in 2^0, 2^1, \ldots$ until it returns a nonempty result.
    The work and depth bounds follow from \Cref{lem:parallel-cover}.
\QED

%% file: additive.tex
\section{Additive Approximation for the Girth}
\label{sec:additive}
As discussed in the introduction, combining Theorem~\ref{thm:girthapx} with the BFS computation of the lengths of shortest cycles through all nodes in a random sample of size $\tilde{O}(n^{1-a})$, yields the following corollary:

\begin{recall}{Corollary~\ref{cor:add}} For all $a\in (0,1)$, there is an $\tilde{O}(mn^{1-a})$ time combinatorial algorithm that w.h.p returns an $O(n^a)$ additive approximation to the girth of an unweighted directed graph.
\end{recall}

It is unclear whether the algorithm from the above corollary can be derandomized.
The algorithm uses randomization in many places: (1) it uses a random sample to hit long cycles that we don't have a handle on otherwise, (2) it uses sampling quite heavily to obtain estimates of the sizes of reachability sets of all vertices, (3) it grows random neighborhoods according to an exponential distribution.

We are not aware of any deterministic approach that achieves running time $O(mn^{1-\eps})$ for $\eps>0$ for any of the above cases. In fact, as far as we know, the only way to achieve (2) deterministically is to compute the reachability trees explicitly.
Despite this, we show that the result can be partially derandomized
 using different techniques:

\begin{recall}{Theorem~\ref{thm:additve-appx}}
Let $G=(V,E)$ be a directed unweighted graph with unknown girth $g$, and let $0<a,\eps< 1$ be parameters. There is a {\em deterministic} combinatorial algorithm that computes in $\tilde{O}((1/\eps^2)mn^{1-a})$ time a cycle whose length is
\begin{enumerate}
    \item an $O(n^{a})$ additive approximation of $g$ if $g \le n^a$, and
    \item a $(1+\eps)$ multiplicative approximation of $g$ if $g > n^a$.
\end{enumerate}
\end{recall}

In the the reminder of this section we prove Theorem~\ref{thm:additve-appx}.



Roughly speaking, our algorithm works in iterations, where each iteration takes $\tilde{O}((1/\eps) m)$ time. Let $C$ be a shortest cycle in $G=(V,E)$. The idea of the algorithm is as follows. In each iteration we consider a shortest path of $\lceil \eps \cdot n^a \rceil$ vertices. If no such path exists, then the diameter of $G$ must be smaller than $\eps n^a$, and we can pick any cycle and return it as our approximation.
Assume now that there is a shortest path $P$ with at least $\lceil \eps \cdot n^a \rceil$ vertices. Either $P\cap C \neq \emptyset$, or we can remove $P$ from $G$ and recurse on the remaining graph. If $P\cap C \neq \emptyset$, our algorithm finds an approximation for $C$ by constructing a new weighted graph $G'$ and a shortest path $P'$ between two nodes $s$ and $t$ in $G'$ whose second shortest simple path length is a good approximation to the length of $C$.

If we could compute this second shortest path exactly, then we would be done. Unfortunately, the
fastest known algorithm for second shortest path takes $O(mn + n^2\log\log n)$ time~\cite{GotthilfL09}, and moreover Vassilevska W. and Williams~\cite{WilliamsW10} showed that the problem is subcubically equivalent to APSP, so that a truly subcubic algorithm for it would be a breakthrough. Our goal is to obtain an almost linear time algorithm, however, since we might need to repeat the procedure $n^{1-a}$ times (removing $n^a$ nodes in each iteration).

Fortunately, Bernstein\cite{Bernstein10} developed an $\tilde{O}(m/\eps)$ time algorithm that computes  a $(1+\eps)$ multiplicative approximation for the second shortest simple path in directed weighted graphs. We use this algorithm for our $\tilde{O}(mn^{1-a})$ mixed approximation algorithm.

Before we formally describe our algorithm, we note that a cycle in a directed graph must be contained in a strongly connected component (SCC). We can assume that $G$ is strongly connected, as otherwise we compute in $O(m)$-time its SCCs and run the algorithm in every non-singleton SCC. If all  SCCs are singletons, then the graph is a directed acyclic graph and has no cycles.

We start by taking an arbitrary vertex $z$ of $G$ and using BFS in $O(m)$ to find the longest shortest paths  $Q_{in},Q_{out}$, in and out of $z$, respectively. Let $Q$ be the longer of $Q_{in}$ and $Q_{out}$. By the triangle inequality, $Q$ must have length at least half of the diameter of $G$ (notice that since $G$ is strongly-connected, the diameter is well-defined). Let $d=\lceil \eps \cdot n^a \rceil$. If the length of $Q$ is $<d$, then the diameter is $<2d$, and any vertex of $G$ is on a cycle of length at most $2d$: take the edge $(x,y)$ on a shortest cycle $\tilde{C}$ through $x$; the length of $\tilde{C}$ is $1+d(y,x)\leq 1+(2d-1)=2d.$ Therefore, by running a BFS from an arbitrary vertex and stopping when the first backward edge is detected, we  find a cycle that is an $O(\eps n^a)$ additive approximation to the shortest cycle.

Otherwise, the diameter is at least $2d$. Let $P=\langle v_d,\dots, v_0\rangle$ be a portion of $d$ edges from the path that we have computed. We construct a new directed weighted graph $G'$ as follows.
\begin{enumerate}
\item Initialize $G'$ to be $G$. Set all weights to $1$.
\item Add the following vertices and edges to $G'$.
    \begin{enumerate}
        \item For each $v_i\in P$, where $i \in\{0,\dots,d\}$, create new nodes $u_i$ and $u'_i$.
        \item For each $i \in\{0,\dots,d\}$ add an edge from $u_i$ to $u'_i$, and for each $i \in\{0,\dots,d-1\}$ add an edge from $u'_i$ to $u_{i+1}$. All edges are of  weight $1$.
    \end{enumerate}

\item For each $v_i\in P$, where $i \in\{0,\dots,d\}$, add the following new edges to $G'$.
    \begin{enumerate}
        \item Add a new edge of weight $4d -3i$ from $u_i$ to $v_i$.
        \item For each outgoing edge $(v_i, x)\in E$ of $v_i$, add a new edge of weight $4d -3i$ from $u_i$ to $x$.
        \item For each incoming edge $(y, v_i)\in E$ of $v_i$, add a new edge of weight $3i$ from $y$ to $u'_i$ .
    \end{enumerate}
\end{enumerate}
From the above construction it follows that there is a path $P'=\langle u_0,u'_0,u_1,u'_1, \dots, u_d, u'_{d}\rangle$ in $G'$ of length $2d+1$.
Moreover, $P'$ is the shortest path from $u_0$ to $u'_d$. To see this, notice first that there is no edge from $u$ to $v$, where $u,v\in P'$ are not consecutive. Therefore, any  path from $u_0$ to $u'_d$ other than $P'$, contains a vertex $v\notin P'$. The  length of such a path is at least $4d>2d+1$ since it must use an edge of weight $4d -3i$ to leave $P'$ and an edge of weight $3j$ to return $P'$, where $0\le i \le j \le d$.

Next we apply Bernstein's algorithm to find a second shortest path for $P'$ in $G'$. Similarly to prior work on replacement paths, given a shortest path $Q$ we say that a path $D(u,v)$ is a \emph{$\langle u,v \rangle$-detour} of $Q$ if $D(u, v)$ is a simple path for which $D(u, v)\cap Q=\{u, v\}$ and $u$ precedes $v$ on $Q$.
It is easy to show that the second shortest path $Q'$ of $Q=\langle v_0,\dots,v_k \rangle$ has the following form: $Q'=Q(v_0, u)\cdot D(u, v) \cdot Q(v, v_k)$, where $Q(v_0, u)$ and  $Q(v, v_k)$ are the subpaths of $Q$ from $v_0$ to $u$ and from $v$ to $v_k$, respectively, and $D(u, v)$ is a \emph{$\langle u,v \rangle$-detour} of $Q$ (e.g. see Lemma 2.1 in~\cite{Bernstein10}).

The following fact follows easily from the construction of $G'$ and $P'$ above.
\begin{fact}\label{fact:detour}
If $P'$ has a \emph{$\langle u_i,u'_j \rangle$-detour} then it has the following structure:
$(u_i, x) \cdot Q' \cdot (y,u'_j)$, where $Q'$ is a path from $x$ to $y$ in $G$, and $x$ is an out-neighbor of $u_i$ in $G$ and $y$ is an in-neighbor of $v_j$ in $G$. Notice $Q'$ might be an empty path.
\end{fact}

In the next lemmas we establish  the relationship between a shortest cycle that intersects $P$ in $G$ and a second shortest path for $P'$ in $G'$.

\begin{lemma}
\label{lemma:detour-cycle}
Let $0\le i \le j \le d$. If  $P'$ has a \emph{$\langle u_i,u'_j \rangle$-detour} $D(u_i,u'_j)$, then there is a cycle in $G$ that contains $P(v_j, v_i)$ and has length $\leq |D(u_i,u'_j)|+|P(v_j, v_i)|=|D(u_i,u'_j)|+ (j-i)$.

Furthermore, if  $G$ has a simple cycle $C$ that contains $P(v_j, v_i)$, then $P'$ has a \emph{$\langle u_i,u'_j \rangle$-detour} of length at most $|C|-|P(v_j, v_i)|+1$.
\end{lemma}
\Proof
Let $Q$ be a \emph{$\langle u_i,u'_j \rangle$-detour} of $P'$. From the construction of $G'$ it follows that
$Q=(u_i, x) \cdot Q' \cdot (y,u'_j)$
where $Q'$ is a path from $x$ to $y$ in $G$\footnote{Notice it is possible that $x=y$, and then $\langle v_i, x \rangle \cdot Q' \cdot \langle y, v_j \rangle$ is actually $\langle v_i, x, v_j \rangle$. It is also possible, in addition, $v_i = x=y$, and then $\langle v_i, x \rangle \cdot Q' \cdot \langle y, v_j \rangle$ is actually $\langle v_i, v_j \rangle$; this is the reason for adding the edges $(u_i,v_i)$ in $G'$. For simplicity of the presentation, we assume the concatenation notation subsume all these cases.}.

We show that $C = ( v_i, x ) \cdot Q' \cdot ( y, v_j ) \cdot P(v_j, v_i)$ is a cycle in $G$.
From the construction of $G'$ it follows that the edges $(u_i, x)$ and $(y, u'_j)$ in $G'$ correspond to the edges $(v_i,x)$ and $(y,v_j)$ in $G$, respectively, and since the path $Q'$ is also in $G$ it follows that $C$ is a cycle in $G$.



Let $C$ be a simple cycle such that $P(v_j, v_i)\subseteq C$ for some $i,j$ (possibly equal). If $C = \langle v_j,\dots,v_i \rangle$ (i.e. $(v_i,v_j)\in E$), then $i\neq j$ and we have the following \emph{$\langle u_i,u'_j \rangle$-detour}:
$D(u_i, u'_j)= \langle u_i, v_i, u'_j \rangle$. Otherwise, we have a shortest path $B$ from $v_i$ to $v_j$, such that $B\cap P = \{v_i, v_j\}$ and $B \not= \{v_i, v_j\}$. Let $x$ be the vertex that follows $v_i$ in $B$ and $y$ the vertex the precedes $v_j$ in $B$ (it might be that $x=y$), then we have the following  \emph{$\langle u_i,u'_j \rangle$-detour}:
$D(u_i, u'_j)= ( u_i, x ) \cdot B(x,y) \cdot ( y, u'_j )$.
\QED

\begin{lemma}
\label{lemma:ssp-length-bound}
Let $C^*$ be a shortest cycle in $G$. If $P(v_j, v_i)\subseteq C^*$, where $j\ge i$, then the length of a second shortest path of $P'$ is at most $6d - 2 + |C^*|$.
\end{lemma}
\Proof
It follows from Lemma~\ref{lemma:detour-cycle} that $P'$ has a \emph{$\langle u_i,u'_j \rangle$-detour}. From Fact~\ref{fact:detour} it has the following structure:
$(u_i,x)\cdot Q'(x, y) \cdot (y,u'_j)$.
Consider the path $P'(u_0, u_i)\cdot (u_i,x)\cdot Q'(x, y) \cdot (y,u'_j) \cdot P'(u'_j, u'_d)$.
Its length is $2i + (4d - 3i) +  d_G(x,y) + 3j + 2(d-j) = 6d  + (j - i) +  d_G(x,y) = 6d - 2 + d_G(x,y) + 2 + (j - i) \le 6d - 2 + d_G(v_i,v_j) + d_G(v_j,v_i) = 6d - 2 + |C^*|$.
%
%
\QED

According to Lemma~\ref{lemma:detour-cycle}, a second shortest path implies a cycle $C$ in $G$ consisting of the detour of the second shortest path together with the path in $G$ corresponding to the subpath of $P'$ that was circumvented. Notice it is easy to derive from $C$ a simple cycle in $G$, which might be shorter. Denote by $L$ the length of a second shortest path. The length of $C$ is then at most $d+L$.

Let $C^*$ be a shortest cycle in $G$. If $P\cap C^* \neq \emptyset$, according to Lemma~\ref{lemma:ssp-length-bound}, $L\le 6d - 2 + |C^*|$. Since we are using a $(1+\eps)$-approximation for the second shortest path ($\eps < 1$), we get a cycle of length at most:
$$d+(1+\eps)L=d+(1+\eps)(6d - 2 + |C^*|) = 7d - 2 + \eps (6d -2) +  (1+\eps)|C^*|\leq O(d) + (1+\eps)|C^*|.$$

If $g\le n^a$, then we found a cycle of size $O(n^a)$. If $g>n^a$, then since $d\leq O(\eps n^a)$, we have a $1+O(\eps)$ multiplicative approximation for the girth.

Thus, if an approximate second shortest path of length $\leq 16d$ is found, we can conclude that $L\leq 16d$ and hence $|C^{*}|\leq O(d)$, so we can stop and return.
Otherwise, we can conclude that none of the vertices of $P$ are on cycles of length $\leq d$ in $G$, as otherwise the algorithm would return an approximate second shortest path of length $7d - 2 + \eps (6d -2) +  (1+\eps)|C^*|< 13d-4+2d < 16d$.
We can thus remove all the vertices of $P$ from $G$ and repeat the process above on the new graph.

Consider the first iteration in which $P'$ contains a vertex of $C^*$.
Since up to this iteration no vertices of $P'$ are removed, the graph will contain a detour corresponding to the portion of $C^*$ not on $P$, and the
approximate cycle returned will be of length $\leq O(d) + (1+\eps)|C^*|,$ as argued above.
If the girth is $\leq 2d$, the approximation is additive $O(d)$, and otherwise it is multiplicative $1+O(\eps)$.


The correctness of the algorithm follows from the discussion above. The runtime is as follows. The time to decompose the graph to its strongly-connected components is $O(m+n)$~\cite{Tarjan72}.
The time to construct $G'$ is $O(m)$, and the running time of Bernstein's algorithm for the second shortest path on $G'$ is $\tilde{O}(m / \eps)$. We conclude that the running time for a single iteration is $\tilde{O}(m / \eps)$. The number of iterations we have in the algorithm is at most $\lceil (1/\eps) n^{1-a} \rceil$, since we are removing $\lceil \eps \cdot n^a \rceil$ vertices from $G$ in each iteration. It follows that the total running time of the algorithm is $\tilde{O}((1/\eps^2) m n^{1-a})$, thus proving Theorem~\ref{thm:additve-appx}.

%% file: lower.tex
\section{Lower Bounds}
\label{sec:lower}
In this section we provide a conditional lower bound for the problem of computing additive approximations for the girth of a directed unweighted graph.

Let us begin with our plausible hypothesis:

\begin{hypothesis}\label{hyp-triangle-detection}
Any combinatorial (possibly randomized) algorithm for triangle detection in $n$-node $m$-edge graphs with $m=\Theta(n^2)$ requires (expected) $n^{3-o(1)}$ time.
\end{hypothesis}

{\em Combinatorial} algorithms informally do not use Strassen-like matrix multiplication, and hopefully do not hide high constants in the big-O. The current best combinatorial algorithms for triangle detection run in time $\min\{O(n^3/\log^4 n), O(m^{3/2})\}$ time~\cite{Yu15,ItaiR78}. It is a major open problem to design a truly subcubic, i.e. an $O(n^{3-\eps})$ time combinatorial algorithm for constant $\eps>0$ for triangle detection. Triangle detection is known~\cite{WilliamsW10} to be {\em subcubically equivalent} to Boolean Matrix Multiplication (BMM) under combinatorial fine-grained reductions, and thus the above hypothesis is equivalent to the hypothesis that combinatorial BMM of $n\times n$ matrices requires $n^{3-o(1)}$ time.

We now state our result:
\begin{theorem}\label{lb:add}
Under Hypothesis~\ref{hyp-triangle-detection}, any combinatorial algorithm that computes an additive $n^{1/2}-1$ approximation to the girth of all directed $n$-node, $m=O(n)$-edge graphs requires $mn^{1/2-o(1)}$ time.
%
%
%
\end{theorem}

%
%
%
%
%

\Proof{
Let $G=(V,E)$ be an $n$-node, $m$-edge directed graph for $m=\Theta(n^{2})$, so that we want to detect the presence of a $3$-cycle in $G$.
We now create a new directed $H$ as follows:
\begin{itemize}
\item $H$ has $n^{2}$ vertices: for every $v\in V$ we create $n$ copies $v_1,\ldots,v_{n}$.
\item For every edge $(u,v)\in E$ of $G$, we create directed edges $(u_{n},v_{1})$ and $(u_{i},v_{i+1})$ for all $i\in \{1,2\}$.
\item For every vertex $v\in V$, create directed edges $(v_i,v_{i+1})$ for all $i\in\{3,\ldots,n-1\}$.
\end{itemize}

Every triangle $a^1\rightarrow a^2 \rightarrow a^3\rightarrow a^1$ in $G$ is represented by an $n$-cycle in $H$: $a^1_{n}\rightarrow a^2_1\rightarrow a^3_2\rightarrow a^1_3\rightarrow a^1_{4}\rightarrow\ldots\rightarrow a^1_{n}$.


Every $n$-cycle in $H$ must correspond to a $3$-cycle in $G$, as there is a path from $v_3$ to $w_{n}$ if and only if $v=w$. Moreover, any cycle in $H$ has length that is a multiple of $n$ as each cycle must go through all $n$ partitions of the graph over and over until it lands at the same node.
The girth of $H$ is thus either $n$ if $G$ has a $3$-cycle, or at least $2n$ otherwise.
$H$ has $N=n^{2}$ vertices and $M\leq 3m + n^{2}=\Theta(n^{2})$ edges.

Suppose that there is some constant $\eps>0$ such that for all $a$ there is an $O(MN^{1/2-\eps})$ time algorithm that achieves an additive $N^{1/2}-1$ approximation to the girth of $M$-edge, $N$-node directed graphs.
Let's apply this algorithm to $H$. If it finds an additive $N^{1/2}-1 = n-1$ -approximation to the girth of $H$, it will be able to detect whether $G$ contains a triangle. The running time of the algorithm would be
$$O(MN^{1/2-\eps}) = O(n^2\cdot n^{1-2\eps})=O(n^{3-2\eps}),$$
which contradicts Hypothesis 1.
\QED

Considering multiplicative approximation for the girth in directed unweighted graphs, it is known that any truly subcubic combinatorial algorithm that computes a $2-\eps$ approximation ($0<\eps<1$) for the girth in directed unweighted graphs, implies a truly subcubic time combinatorial algorithm for triangle detection. 
This is formalized in the next probably folklore theorem. A formal proof of it appears in~\cite{RodittyW12}.
\begin{theorem}[Folklore]
Under Hypothesis~\ref{hyp-triangle-detection}, any combinatorial algorithm that for $\epsilon \in (0,1)$ computes a multiplicative $2-\eps$ approximation for the girth of a directed $n$-node, $m$-edge graph requires $n^{3-o(1)}$ time.
\end{theorem}

%
%
}

%% file: ball-app.tex
\section{Ball Size Estimation}
\label{sec:ball_size_estimation}
	
Here we present a routine that can estimate the sizes of neighborhoods of all vertices. The approach is similar to that of~\cite{Cohen97}. We provide the algorithm, $\textsc{Estimate-Balls}$, that when invoked on a graph $G$ with radius parameter $r$ and error parameter $\epsilon$ computes w.h.p. for every vertex the fraction of vertices that it can reach at distance at $r$ and the fraction of vertices that can reach it.

\begin{figure}[ht]
	\noindent
	\centering
	\fbox{
		\begin{minipage}{6in}
            \noindent $(s^{out}, s^{in}) = \textsc{Estimate-Balls} (G, r, \epsilon)$, where $G = (V, E, l)$ is a directed graph and $r, \epsilon > 0$.
			\begin{enumerate}
                \item Sample $t = \lceil 20\cdot \log n / \epsilon^2 \rceil$ vertices $v_1, \ldots, v_t$  independently uniformly at random from $V$ with replacement.
                \item Compute the distances between every vertex in $V$ and each of $v_1, \ldots, v_t$.
                \item For each $u \in V$, let $s^{out}_u$ be the fraction of $v_1, \ldots, v_t$ such that $d_G(u, v_i) \leq r$.
                \item For each $u \in V$, let $s^{in}_u$ be the fraction of $v_1, \ldots, v_t$ such that $d_G(v_i, u) \leq r$.
                \item Return $(s^{out}, s^{in})$.
			\end{enumerate}
		\end{minipage}
	}
	\caption{The algorithm for estimating the sizes of out- and inballs at radius $r$ for a given graph.}
	\label{fig:estimate}
\end{figure}

Our algorithm simply samples vertices with replacement and computes distances to and from them to estimate the ball sizes. The analysis of $\textsc{Estimate-Balls}$ reduces to a simple application of Chernoff bounds and union bound. We prove that it works in Lemma~\ref{lem:balls}. 

\begin{lemma}
\label{lem:balls}
    Let $s^{out}, s^{in}$ be the output of $\textsc{Estimate-Balls}(G, r, \epsilon)$.
    For any vertex $u$, let $\bar{s}^{out}_u$ be the fraction of vertices in $V$ such that $d_G(u, v_i) \leq r$.
    Then, whp., for all vertices $u$ it holds that $|\bar{s}^{out}_u - s^{out}_u| \leq \epsilon$.
    An analogous statement holds for $s^{in}$. The algorithm runs in time $O(m \epsilon^{-2} \log^2 n)$.
\end{lemma}
\Proof
    By a standard Chernoff bound, we have
    \[
        \Pr[|\bar{s}^{out}_u - s^{out}_u| > \epsilon] \leq 2\exp(-2t\epsilon^2)
                                                    \leq 2\exp(-40 \log n)
                                                   = 2n^{-40} ~.
    \]
    An analogous bound holds for $s^{in}$.
\QED

%% file: sequential_clustering.tex
\section{Exponential Distributions}
\label{sec:gen_tools}

Here we recall some basic facts about the exponential distribution we use in the paper.

\begin{lemma}[Exponential Distribution Facts]
	\label{lem:exp-dist}  We let $\Exp(\alpha)$ denote the exponential
	distribution with parameter $\alpha$. This distribution is supported
	on $\R_{\geq0}$ with a pdf given $p(x)=\alpha\cdot\exp(-\alpha x)$.
	This distribution has the following properties:
	\begin{itemize}
		\item \textbf{CDF}: $\Pr[\Exp(\alpha)\leq x]=1-\exp(-\alpha x)$ for $x\geq0$.
		\item \textbf{Expected Value}: $\E\Exp(\alpha)=\frac{1}{\alpha}$ 
		\item \textbf{Memoryless}: $\Pr\left[\Exp(\alpha)\geq s+t\,|\,\Exp(\alpha)\geq s\right]=\Pr\left[\Exp(\alpha)\geq t\right]$
        \item \textbf{High Probability}: The maximum of $n$ independent r.v.s drawn from $\Exp(\alpha)$ is $O(\frac{\log n}{\alpha})$ with high probability.
	\end{itemize}
\end{lemma}

\Proof
Direct calculation reveals that
\[
\Pr\left[\Exp(\alpha)\leq x\right]=\int_{-\infty}^{x}\alpha\exp(-\alpha x)=-\exp(-\alpha x)+\exp(0)=1-\exp(-\alpha x)
\]
giving the formula for the CDF. Furthermore, integration by parts
yields that 
\begin{align*}
	\E\Exp(\alpha) & =\int_{0}^{\infty}\alpha x\exp(-\alpha x)dx=\left[-x\exp(-\alpha x)\right]|_{0}^{\infty}-\int_{0}^{\infty}-\exp(-\alpha x)dx\\
	& =-\frac{1}{\alpha}\exp(-\alpha\infty)+\frac{1}{\alpha}\exp(-\alpha0)=\frac{1}{\alpha}
\end{align*}
giving the expected value formula. Direct calculation again yields
\begin{align*}
	\Pr\left[\Exp(\alpha)\geq s+t\,|\,\Exp\geq s\right] & =\frac{\Pr[\exp(\alpha)\geq s+t]}{\Pr\left[\exp(\alpha)\geq s\right]}=\frac{\exp(-\alpha(s+t))}{\exp(-\alpha s)}\\
	& =\exp\left(-\alpha t\right)=\Pr\left[\Exp(\alpha)\geq t\right]
\end{align*}
proving the memoryless property. Finally the high probability bound
is immediate from the CDF and the definition of high probability.
\QED

\section{Sequential Clustering Algorithm}
\label{sec:sequential}

Here we formalize and prove the alternative approach to clustering described in \Cref{sec:approach:spanner}.

\begin{figure}[ht]
	\noindent
	\centering
	\fbox{
		\begin{minipage}{6in}
            \noindent $(V_1, V_2, \ldots) = \textsc{Sequential-Cluster-Out(-In)} (G, (v_1, \ldots, v_{t - 1}), r)$, where $G = (V, E, l)$ is a directed graph, $v_1, \ldots v_{t -1} \in V$, and $r> 0$.
			\begin{enumerate}
                \item Set $\beta := \log (n) / r$.
                \item Let $G_0 := G$.
                \item For $i$ in $1, \ldots, t$:
                    \begin{enumerate}
                        \item If $v_i$ is not in $G_{i-1}$, let $G_i := G_{i-1}, V_i := \emptyset$ and continue the loop. Otherwise:
                        \item Pick $x_i \sim \Exp(\beta)$.
                        \item Let $V_i := \outball_{G_{i-1}}(v_i, x_i)$. ($\inball_{G_{i-1}}(v_i, x_i)$ for $\textsc{Cluster-In}$)
                        \item Let $G_i := G_{i-1}$ with $V_i$ and incident edges removed.
                    \end{enumerate}
                \item Return $(V_1, V_2, \ldots, V_{t-1}, V \setminus \bigcup_i V_i)$
			\end{enumerate}
		\end{minipage}
	}
	\caption{The sequential clustering algorithm.}
	\label{fig:cluster}
\end{figure}

\begin{lemma}
\label{lem:seqcluster}
    Let $(V_1, V_2, \ldots, V_t) = \textsc{Sequential-Cluster-Out}(G, (v_1, \ldots, v_t), r_0, r)$ (analogously of $\textsc{Sequential-Cluster-In}$). Then for any $c\geq 1$ we have
    \begin{enumerate}
        \item with probability at least $1 - n^{1 -c}$ for all $i < t$, the radius of the tree corresponding to $V_i$ is at most $c \cdot r$, whp., 
        \item for any pair of vertices $u, v$ at roundtrip distance at most $R$ in $G$, they are in the same set $V_i$ with probability at least $\exp(-\log (n) R / r)$.
    \end{enumerate}
\end{lemma}
\Proof
Note that the maximum radius of any constructed tree is upper bounded by $\max_i x_i$.
For every $i \in 1, \ldots, t$, we have
\[
    Pr\left[x_i \geq c \cdot r\right] \leq \exp(-c \cdot \beta r)
                                        = n^{-c},
\]
and so by union bound the maximum radius is at most $c \cdot r$ with probability at least $1 - n^{1 - c}$.

To prove the
remainder of the lemma, fix two vertices $u$ and $v$ at roundtrip distance at most $R$ in $G$.
Assume the $i$-th cluster is the first one to contain an element of the set $\{u, v\}$.
By the memoryless property of the exponential distribution we see that conditioned on this event
the probability that cluster $i$ contains both vertices $u$ and $v$ is at least
\begin{align*}
    \Pr\left[\Exp(\beta)\geq R\right] = \exp(-\beta R),
\end{align*}
yielding the desired result.
\QED